\documentclass[a4paper,11pt,hyper]{JHEP3}
\usepackage{amsfonts,latexsym,epsfig,amssymb,amsmath,mathrsfs,ulem,cite}

\pdfoutput=1

\newcommand{\ogw}{\Omega_{\rm GW}}

\newcommand{\agt}{\hspace{0.3em}\raisebox{0.4ex}{$>$}\hspace{-0.75em}\raisebox{-.7ex}{$\sim$}\hspace{0.3em}}
\newcommand{\bm}[1]{\hbox{\boldmath{$#1$}}}
\newcommand{\sbm}[1]{\hbox{\boldmath{\scriptsize$#1$}}}

\usepackage{subfigure}

\title{Gravitational wave forest from string axiverse}
\author{
Naoya Kitajima$^a$, Jiro Soda$^b$, Yuko Urakawa$^{a, c}$\\
a.~Department of Physics and Astrophysics, Nagoya University, Chikusa,
Nagoya 464-8602, Japan\\
b.~Department of Physics, Kobe university, Kobe 657-8501, Japan\\
c.~Institut de Ciencies del Cosmos,
Universitat de Barcelona, Marti i Franques 1 08028, Barcelona, Spain
}

\abstract{
Axions predicted in string theory may have a
scalar potential which has a much shallower potential region than the
conventional cosine potential. We first show that axions which were
located at such shallow potential regions generically undergo prominent
resonance instabilities: the well-known narrow resonance and/or the
flapping resonance, which has not been well investigated. 
We also study non-linear dynamics of axions caused by these resonance
instabilities based on lattice simulation. We find that string axions in
various mass ranges generate gravitational waves (GWs) with peaks at various frequencies determined by the mass
scales, dubbed the GW forest. This may allow us to explore string axiverse
through future multi-frequency GW observations. We also investigate GWs produced by the axion which accounts for present dark matter component.
}

\preprint{KOBE-COSMO-18-07}

\maketitle
\begin{document}

\section{Introduction}

After the first direct detection of gravitational waves
(GWs)~\cite{Abbott:2016blz}, we are now in an era of GWs.
An important lesson from the history of astronomy with electromagnetic
waves is that multi-frequency observations of GWs become important.
Roughly speaking, there are three categories of GW sources: astrophysical, cosmological, and primordial sources.
GWs from astrophysical sources in the kHz band  have been detected by GW
interferometers. It is now compelling to discover the primordial GWs
from inflation by measuring B-mode polarization of cosmic microwave
background~\cite{Ade:2017uvt, Matsumura:2013aja}. In addition to these sources, there are several cosmological GW
production processes. For example, parametric amplification of scalar
fields during reheating after inflation, cosmic strings, bubbles
produced at the first order phase transition belong to this
category. As a preparation for the forthcoming era of multi-frequency GW observations by e.g. space-based~\cite{GWspace} and ground-based detectors~\cite{GWground} and pulsar timing observations~\cite{PTA}, 
it is worth exploring other cosmological sources of GWs with a wide
range of frequencies.

From a theoretical viewpoint, the GWs from astrophysical sources have brought us the information on
strong gravity, black hole physics, nuclear physics, star formation, and
so on. A detection of the primordial GWs, which will be a first evidence
of quantized gravitons, will tell us the energy scale of inflation. GWs
from bubbles and cosmic strings provide us the information on
cosmological phase transitions. Along this line, it is natural to ask if
we can probe fundamental physics through GW observations.

String theory predicts the presence of
six dimensional internal space in addition to our four dimensional world.  
The extra dimensions must be compactified and the compactification of the internal space yields various moduli
fields, including axion fields in the four dimensional low energy effective field theory~\cite{Svrcek:2006yi,LVS}. 
Intriguingly, according to string theory, the mass spectrum of axions can be logarithmically flat. 
Hence the axions with various masses are ubiquitous in the Universe, 
 which is named string axiverse~\cite{StAx}. It is, therefore, worth
 exploring imprints of string axiverse in the history of the Universe by
 means of GWs.

In the previous paper~\cite{JY17}, 
two of the authors initiated a study on GW production from string
axiverse to probe the string compactification. It has been suggested
that the scalar potential of a string axion may be much shallower than the
conventional cosine potential away from the potential
minimum~\cite{Dubovsky:2011tu, NWY}. When it was located at the shallow potential region
before it commences to oscillate, the succeeding dynamics becomes
drastically different from the one for an axion with the cosine
potential. In fact, such an axion generically undergoes various
instabilities after the onset of oscillation. In Ref.~\cite{JY17}, it
was suggested that these instabilities can lead to a detectable emission
of GWs. In this paper, we will investigate this possibility in more
detail based on lattice simulation.

The frequency of the emitted GWs is determined by the mass scale of the
axion. Historically, an emission of GWs triggered by a coherently oscillating scalar field
has been vastly studied in the context of reheating after inflation~\cite{KLS_L,KLS,KT97,Dani17}. 
In this case, frequencies of emitted GWs are typically around 1
GHz~\cite{GarciaBellido:2007af}. In contrast, string axions with various
mass scales lead to GW emissions with various frequency range.
These GWs can be a target of multi-frequency GW observations, providing
a new window to probe string axiverse, dubbed GW forest.

The paper is organized as follows. In section 2, first we briefly
summarize the motivations to consider an axion with a plateau potential,
clarifying our definition of the plateau. Then, we discuss different
types of instabilities based on linear analysis. In section 3, we
investigate the nonlinear dynamics of the axion with lattice
simulation, showing that as a consequence of efficient resonance
instabilities, clumps of oscillating scalar fields, so-called oscillons,
are formed. In section 4, we study that resonance instabilities lead to
a copious emission of GWs. We also study the GWs from the axion which
survives until now as dark matter. The final section is devoted to the
conclusion.

\section{Axion dynamics with plateau potential}

In this section, we discuss the dynamics of an axion which was initially
located at a plateau region in the scalar potential. In this case, the
time evolution of the axion becomes drastically different from the
one for axions with the conventional cosine potential. We show that after
the axion starts to oscillate, various instabilities set in, clarifying qualitative differences from
axions with the cosine potential. These instabilities lead to the highly
inhomogeneous spatial distribution of the axion field even if the axion was initially almost homogeneous.

\subsection{Axion potential with a plateau region}
Axions are conventionally assumed to have the cosine-type potential,
given by 
\begin{eqnarray}
V(\phi) = \Lambda^4 \left[  1 -  \cos\left(\frac{\phi}{f}\right)    \right] \ ,
\end{eqnarray}
where $f$ is the axion decay constant and $\Lambda$ is the dynamical
scale. The axion mass $m$ is given by $m= \Lambda^2/f$. This
potential was derived under the dilute instanton gas
approximation. 

The dilute gas approximation can be broken down, e.g., when the
axion is coupled with pure Yang-Mills gauge fields which are strongly coupled. Witten pointed out this
possibility by considering an SU($N$) gauge theory in large $N$
limit~\cite{Witten:1979vv, Witten:1980sp}. More recently, the scalar
potential of an axion which interacts with an SU($N$) gauge
field in the large $N$ limit was considered in the context of axion inflation in
Ref.~\cite{Yonekura:2014oja}. The axion inflation model where the
potential of the axion was generated through the dynamics of a pure
Yang-Mills gauge theory, was dubbed as {\it Pure natural
inflation}~\cite{NWY}. There, the potential possesses the multi-branch
structure, where different branches are obtained by changing $\phi$ to  
$\phi + 2 \pi n f$ with an integer $n$. As was argued in
Refs.~\cite{NWY, NY17}, the potential of $\phi$ in a single branch is
typically given in the form:
\begin{align}
 & V(\phi) = M^4 \left[ 1 - \frac{1}{(1 + (\phi/F)^2)^\beta} \right] \qquad
 \quad (\beta > 0)  \label{purenatural}
\end{align}   
with $M \sim N \Lambda$ and $F \propto N f$. For $\phi/F \ll 1$, this
potential is well approximated by the quadratic potential, while for
$\phi/F \agt 1$, the potential becomes more flatten and asymptotes to
$M^4$ (there is a caveat for the potential analysis in this
region~\cite{NWY, NY17}). The potential in the single branch does not
preserve the symmetry under $\phi \to \phi + 2 \pi f$, while the
periodic symmetry is recovered for the true vacuum energy, which is 
determined by the minimum values among the different
branches~\cite{Yonekura:2014oja, NWY, NY17}. As shown in
Refs.~\cite{NWY, NY17}, for small values of $F$, the prediction in pure
natural inflation becomes compatible with the constraint from Planck
15~\cite{Planck15}. (See also Refs.~\cite{Shiu:2018wzf, Shiu:2018unx}.)

In general, the scalar potential of an axion predicted in string theory
acquires multiple cosine terms through non-perturbative
effects. A linear combination of several cosine terms can exhibit a
wider shallow potential region than the one for a single cosine
potential as is discussed, e.g., in multi-natural inflation model \cite{Czerny:2014wza,Czerny:2014xja}.

Alternatively, in case the axion has a non-canonical kinetic term or the
axion is non-minimally coupled with gravity, the scalar potential can
get flatten after the canonical normalization. Therefore, even if the
potential of the axion is given by the conventional cosine form, the
scalar potential can have a plateau region after the canonical
normalization. In Refs.~\cite{Kallosh:2013hoa, Kallosh:2013yoa,
Kallosh:2013tua}, it was shown that models of this class, 
dubbed as $\alpha$-attractors, rather generically have similar
potential structures such as 
$f(\tanh \phi/\sqrt{6 \alpha})$ (see also
Ref.~\cite{Ferrara:2013rsa} for an earlier study). These potentials typically exhibit flat plateaus for 
$\phi > \sqrt{6 \alpha}$. Notice that in these arguments, the field
$\phi$ appears as a real part of a modulus field and is not an axion.

Along this line, we investigate the dynamics of an
axion with a shallow region in the scalar potential and phenomenological
consequences. In this paper, we neglect contributions of the axion to
the geometry, focusing on the case where the axion starts to oscillate when
it was a subdominant component of the universe. Then, introducing a normalized
potential $\tilde{V}(\tilde{\phi})$ defined as
\begin{eqnarray}
V(\phi) = (m f)^2\, \tilde{V}( \tilde{\phi})\,, \qquad \tilde{\phi}
 \equiv \frac{\phi}{f},
\end{eqnarray}
we can express the Klein-Gordon equation in a spatially flat FRW universe as
\begin{align}
 & \frac{d^2}{d \tilde{t}^2}  \tilde{\phi} + 3 \frac{H}{m}\, \frac{d}{d
 \tilde{t}} \tilde{\phi} + \frac{\partial^2_{\tilde{\sbm{x}}}}{a^2}
 \tilde{\phi} + \tilde{V}_{\tilde{\phi}} = 0\,, \label{KG}
\end{align}
with $\tilde{t} \equiv mt$ and
$\tilde{\bm{x}} = m \bm{x}$.  Here, $t$ is the cosmic time, $H$ is
the Hubble parameter, and $\tilde{V}_{\tilde{\phi}} \equiv d \tilde{V}/ d \tilde{\phi}$. 
The parameter $f$ corresponds to the decay constant for the conventional cosine
potential with $\tilde{V} = 1 - \cos \tilde{\phi}$. For $a \propto t^p$, we can express the Hubble parameter normalized by $m$ as $H/m= p/\tilde{t}$.
Notice that the axion's dynamics in a fixed geometry
becomes scale free, being independent of the parameters $f$ and $m$. In
contrast, when the axion dominates the universe and the geometry is
determined by the axion, the axion's dynamics
becomes rather different, depending on $f/M_{\rm P}$~\cite{Amin17}, 
where $M_{\rm P}$ denotes the Planck scale.

\subsection{Plateau potential and delayed oscillation}  \label{SSec:Delay}

Next, we consider the time evolution of the background
homogeneous mode. Likewise a time evolution of an inflaton, at early
times when the Hubble friction is large enough, the axion slowly rolls
down the potential, behaving as a cosmological constant. When the
Hubble parameter decreases to a certain value, $H_{\rm osc}$, the axion
starts to oscillate. For the quadratic potential $V= (m \phi)^2/2$, the
Hubble parameter at the onset of the oscillation is given by 
$H_{\rm osc} \simeq m$. Meanwhile, for a scalar potential which is
shallower than the quadratic form, $H_{\rm osc}$ becomes smaller than
$m$.

To evaluate $H_{\rm osc}/m$, in general, we need a numerical
analysis. However, we can understand that the
onset of the oscillation indeed delays for a shallower potential also from a heuristic argument. In the
absence of the cosmic expansion, the Klein-Gordon equation for the
homogeneous mode is given by 
$d^2 \tilde{\phi}/(d \tilde{t}^2) +\tilde{V}_{\tilde{\phi}} =0$ and
the time scale of the potential driven motion is roughly estimated as 
$\tilde{t} = mt \simeq \sqrt{|\tilde{\phi}/\tilde{V}_{\tilde{\phi}}|}$. 
In an expanding universe, the axion commences to oscillate, roughly when the time scale of the
potential driven motion becomes comparable to the one of the cosmic
expansion, $m/H$. Therefore, when the axion was initially situated at a potential region
with $|\tilde{V}_{\tilde{\phi}}/ \tilde{\phi}| < 1$, i.e., at a potential region
which is shallower than $\tilde{\phi}^2$, the oscillation does not start yet
at $H \simeq m$ and $H_{\rm osc}/m$ becomes
smaller than 1, indicating a delayed commencement of the oscillation.

In this paper, to address the situations discussed in the previous
subsection, we consider a scalar potential which satisfies
\begin{itemize}
\item[i)]
$\tilde{V}(\tilde{\phi}) \to \tilde{\phi}^2/2$ in the limit
$\tilde{\phi} \to 0$, 

\item[ ii)] $|\tilde{V}_{\tilde{\phi}}/\tilde{\phi}| \ll 1 $ for $|\tilde{\phi}| > 1$. 
\end{itemize}
By a plateau region, we mean a potential region which satisfies the
second condition. Most of the properties which will be discussed in this paper follow only
from these two conditions. Hence, our discussion can apply to a general scalar
field whose potential fulfills the above two conditions, not only to an axion. When we consider axions,
we additionally require $Z_2$ symmetry of the scalar potential, since
axions are pseudo scalar fields. When the axion was initially located at a
plateau region, since the initial velocity decays immediately and
$\tilde{\phi}$ stays almost constant until the onset of the oscillation,
evaluating $\tilde{V}_{\tilde{\phi}}/\tilde{\phi}$ with an initial value
$\tilde{\phi}_i$, we obtain 
\begin{align}
 & \frac{H_{\rm osc}}{m} \simeq \sqrt{\left|
 \frac{\tilde{V}_{\tilde{\phi}}(\tilde{\phi}_i)}{\tilde{\phi}_i}
 \right|} \ll 1 \,,
\end{align}
indicating a significant delay.

Notice that the onset of the oscillation
is delayed, as far as the axion stays in a potential region with
$|\tilde{V}_{\tilde{\phi}}/ \tilde{\phi}| < 1$ sufficiently long in the
time scale of the cosmic expansion, even if $\tilde{V}$ does not have an
extensive plateau region as is required by the condition ii). For
instance, when the axion is initially located around the potential maximum
of the cosine potential $\tilde{V} = 1 - \cos \tilde{\phi}$, i.e.,
at $\tilde{\phi}_i \simeq \pi$, we obtain 
$$
 \frac{\tilde{V}_{\tilde{\phi}}(\tilde{\phi}_i)}{\tilde{\phi}_i} =
 \frac{\sin \tilde{\phi}_i}{\tilde{\phi}_i}   \simeq 0\,. 
$$
Meanwhile, the variation of $\tilde{V}_{\tilde{\phi}}/ \tilde{\phi}$ in
one Hubble time is given by
$$
 \frac{m}{H} \frac{d}{d (mt)}
 \frac{\tilde{V}_{\tilde{\phi}}(\tilde{\phi})}{\tilde{\phi}}
 \bigg|_{\tilde{t}=\tilde{t}_i} \simeq - \frac{1}{\pi} \frac{m}{H_i} \frac{d \tilde{\phi}_i}{d \tilde{t}_i}\,.
$$  
Therefore, in contrast to the plateau potential, for a significant delay to take place in 
the cosine potential, the initial velocity
$d \tilde{\phi}_i /d t_i$ should be tuned to be a small amplitude. (See
Refs.~\cite{Zhang:2017flu, Zhang:2017dpp} for a time evolution of axion dark matter with the cosine
potential under the fine-tuned initial condition.)

In a plateau region which satisfies the condition ii), the curvature of
the potential 
$\tilde{V}_{\tilde{\phi} \tilde{\phi}} \equiv d^2 \tilde{V}/d^2 \tilde{\phi}^2$ takes a negative
value. Meanwhile, in the limit $\tilde{\phi} \to 0$,
$\tilde{V}_{\tilde{\phi} \tilde{\phi}}$ approaches to 1. Therefore, a
potential $\tilde{V}(\tilde{\phi})$ which fulfills the two properties i) and ii) 
generically has inflection points at
 $\tilde{\phi}  = \pm \tilde\phi_c$  with $|\tilde\phi_c| = {\cal O}(1)$. 
When the axion was initially located at the plateau region with
$|\tilde{\phi}| \gg |\tilde{\phi}_c|$, the subsequent time evolution can
be divided into the following phases, depending on the value of the
potential curvature $\tilde{V}_{\tilde{\phi} \tilde{\phi}}$:
\begin{description}
\item[Phase 1.] Rolling down the plateau region with $\tilde{V}_{\tilde{\phi} \tilde{\phi}} < 0$
\item[Phase 2.] Oscillation between $\tilde{V}_{\tilde{\phi}\tilde{\phi}} < 0$ 
and $\tilde{V}_{\tilde{\phi} \tilde{\phi}} > 0$ 
\item[Phase 3.] Oscillation in $ 0 \leq  \tilde{V}_{\tilde{\phi}
	   \tilde{\phi}} \leq 1$
\item[Phase 4.] Harmonic oscillation with $\tilde{V}_{\tilde{\phi}
	   \tilde{\phi}} \simeq 1$
\end{description}
In most of the time during the phase 1, the axion slowly rolls down the
potential. During the phase 2,\, 3,\, and 4, the axion coherently
oscillates and in particular during the phase 2 and 3, the oscillation
is not yet settled down to the harmonic oscillation.

In the phase 1, 2, and 3, different types of instabilities
set in. In the next subsection, we will show that when the onset of the
oscillation is delayed, the instabilities during the phase 2 and 3
persist rather long. In the phase 4, the
self-interaction of the axion ceases to exist and the axion
undergoes the harmonic oscillation, behaving as a dust component.

\begin{figure}[tbh]
\begin{center}
\begin{tabular}{cc}
\includegraphics[width=9cm]{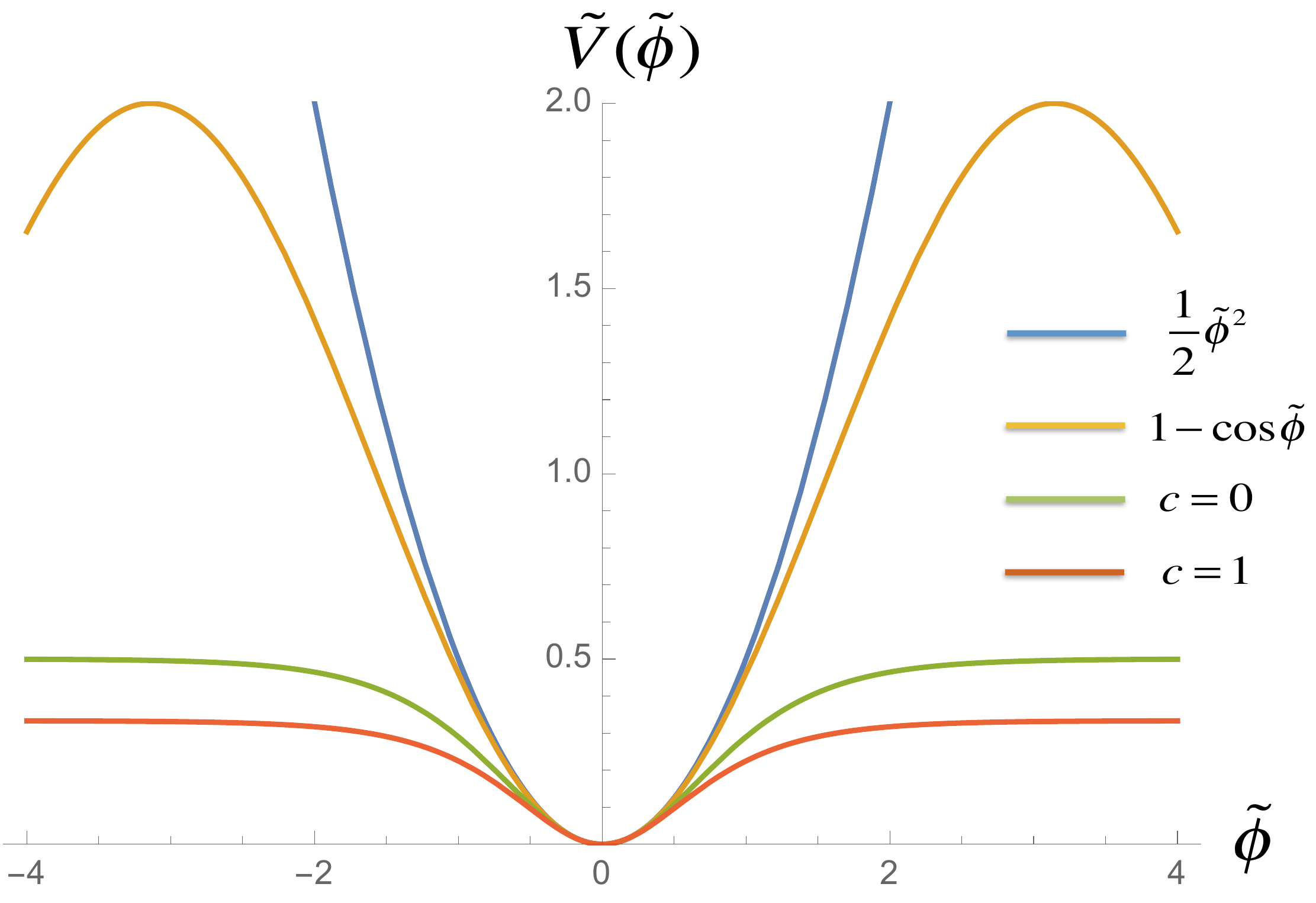}
\end{tabular}
\caption{The plot shows the quadratic potential, the cosine potential,
and the $\alpha$ attractor type potential for $c=0,1$.}
\label{Fg:Vtanhs}
\end{center}
\end{figure}

\begin{figure}[tbh]
\begin{center}
\begin{tabular}{cc}
\includegraphics[width=7cm]{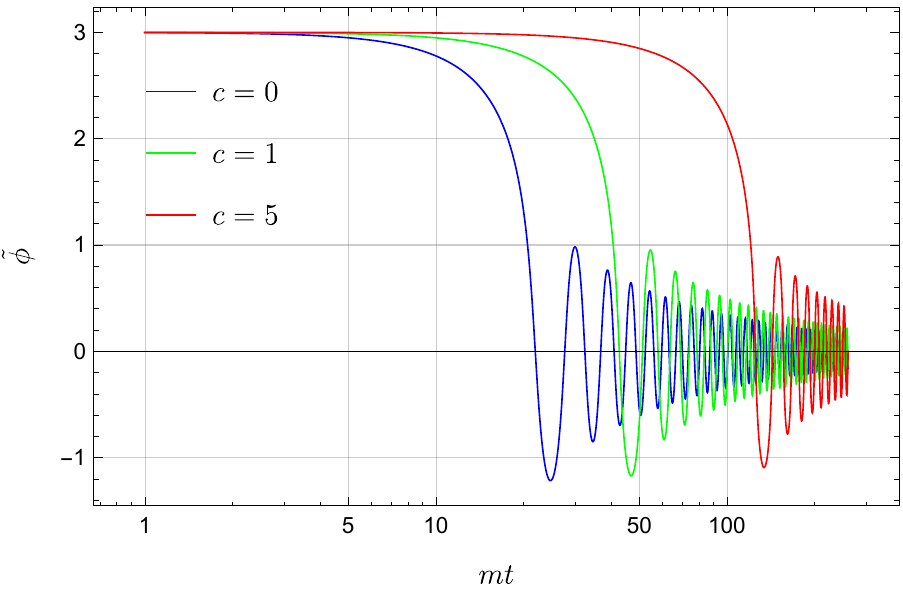}
\includegraphics[width=7cm]{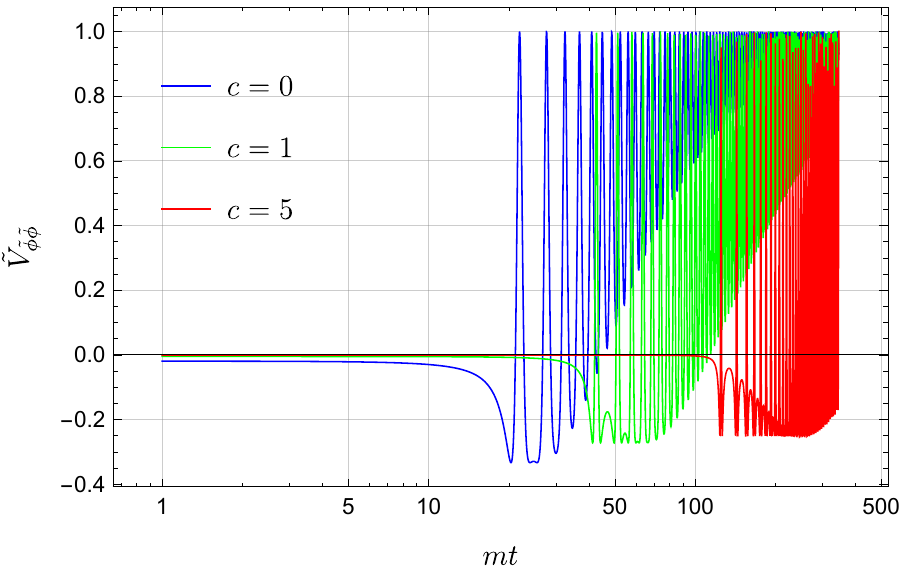}
\end{tabular}
\caption{The left panel shows the time evolution of $\tilde{\phi}$ and
 the right one shows the time evolution of $\tilde{V}_{\tilde{\phi}
 \tilde{\phi}}$ for the $\alpha$ attractor potential. We have taken $\tilde{\phi}_i=3$ and $c=0$ (blue), 1 (green), 5 (red).}
\label{Fg:Vdd}
\end{center}
\end{figure}

As a specific example of the potential which fulfills the above two
conditions, here let us
consider an $\alpha$ attractor type potential~\cite{Kallosh:2013hoa, Kallosh:2013yoa, Kallosh:2013tua} given by  
\begin{eqnarray}
\tilde{V}(\tilde{\phi}) = \frac{1}{2}\frac{(\tanh \tilde{\phi})^2}{1 + c
 (\tanh \tilde{\phi})^{2}}  \label{Exp:Valpha}
\end{eqnarray}
with $c$ being a positive numerical constant. The potential form of Eq.~(\ref{Exp:Valpha}) is shown
in Fig.\ref{Fg:Vtanhs} for $c=0,\,1$ together with the quadratic
potential and the cosine type potential. We numerically solved the background
dynamics for various sets of the parameters
$(c,\,\tilde\phi_i)$ under the slow-roll initial condition during radiation
domination. Fig.\,\ref{Fg:Vdd} shows the time evolution
of $\tilde{\phi}$ (left) and $\tilde{V}_{\tilde{\phi} \tilde{\phi}}$ (right) for 
$(c,\, \tilde{\phi}_i) = (0,\, 3), (1,\, 3), (5,\, 3)$ (the horizontal
axis is $\tilde{t} \equiv mt$). 
The left panel of Fig.\,\ref{Fg:Vdd}
shows the delayed onset of oscillation, i.e. 
$m/H_{\rm osc} = 2 \tilde{t}_{\rm osc} \gg 1$, and it becomes more significant for larger $c$. 
Notice that while the phase 2 finishes after several oscillations for $c=0$,
it continues much longer for $c=5$. 
Since the Hubble friction is no longer important during the
oscillation period, $\tilde{V}_{\tilde{\phi} \tilde{\phi}}$
significantly deviates from 1 over many periods of the oscillation.

\subsection{Instability of inhomogeneous modes}

In this subsection, we study the dynamics of inhomogeneous modes of the axion
during the phase 1, 2, and 3 based on linear analysis. Perturbing
Eq.~(\ref{KG}), we obtain the evolution equation for Fourier modes of the dimensionless
linear perturbation $\delta \tilde{\phi}_k \, (= \delta \phi_k/f)$ as
\begin{align}
\frac{d^2}{ d \tilde{t}^2} \delta \tilde{\phi}_k +  3\, \frac{H}{m}
 \frac{d}{d \tilde{t}} \delta \tilde{\phi}_{k} + \left(\frac{k}{a
 m}\right)^2  \delta \tilde\phi_{k}
 +  \tilde{V}_{\tilde\phi \tilde\phi}\,\delta  \tilde\phi_{k} =0\,, \label{Eq:varphi}
\end{align}
where we neglected the metric perturbations. This can be verified when the
self-interaction of the axion dominates the gravitational interaction,
which is the case of our interest.

For a computational simplicity, let us introduce another variable as
$\varphi_k = a^{3/2} \delta \tilde\phi_k$. The equation of motion
(\ref{Eq:varphi}) is now recast into
\begin{align}
 & \frac{d^2}{d \tilde{t}^2}\varphi_k + \omega^2_k\, \varphi_k = 0\,  \label{Eq:eomchi}
\end{align}
with 
\begin{align}
 & \omega^2_k  \equiv  \left( \frac{k}{a m} \right)^2 +
 \tilde{V}_{\tilde{\phi} \tilde{\phi}}
  -\frac{9}{4} \left( \frac{H}{m} \right)^2 -\frac{3}{2}
 \frac{\dot{H}}{m^2} \simeq  \left( \frac{k}{a m} \right)^2 +
 \tilde{V}_{\tilde{\phi} \tilde{\phi}}\,.  \label{Exp:omega2} 
\end{align}
Here and hereafter, focusing on the case with the delayed oscillation, we
drop the terms which are suppressed for $H/m \ll 1$.

Eq.~(\ref{Eq:eomchi}) can be regarded as one dimensional kinematics
governed by the quadratic potential 
\begin{equation} 
{\cal V}_k \equiv \frac{1}{2} \omega_k^2 \varphi_k^2 \simeq  \frac{1}{2}
\left[ \left(\frac{k}{a m}\right)^2 + \tilde{V}_{\tilde{\phi} \tilde{\phi}} \right] \varphi_k^2. \label{eq:Vk} 
\end{equation}
The first term in the right hand side always acts as a restoring force,
which drives $\varphi_k$ towards the origin. On the other hand, the
potential driven force, the second term, acts more non-trivially, since
it flips the signature at the inflection points. In the previous
subsection, we divided the time evolution of the homogeneous mode into
four different phases, depending on the value of
$\tilde{V}_{\tilde{\phi} \tilde{\phi}}$. In the following, we will 
show that inhomogeneous modes undergo different types of instabilities
in different phases.

\subsubsection{Tachyonic instability: Phase 1}
During the phase 1, (the homogeneous mode of) the axion rolls down the
plateau region, where $\tilde{V}_{\tilde{\phi} \tilde{\phi}}$ takes a
negative value. During this period, $\omega_k^2$
takes a negative value for the low-$k$ modes with
\begin{align}
 & \frac{k}{am} < \sqrt{|\tilde{V}_{\tilde{\phi} \tilde{\phi}}|}\,, \label{Exp:TI}
\end{align}
and these modes grow exponentially. In particular, the growth rate of
the modes with $k/(am) \ll \sqrt{|\tilde{V}_{\tilde{\phi} \tilde{\phi}}|}$ 
become independent of $k$. Such a tachyonic instability occurs also during the reheating
process \cite{Felder:2000hj}.

\subsubsection{Flapping resonance: Phase 2}

In this phase, the axion goes back and forth between the regions with 
$\tilde{V}_{\tilde{\phi} \tilde{\phi}} < 0$ and 
$\tilde{V}_{\tilde{\phi} \tilde{\phi}} > 0$. During the term when
$\tilde{V}_{\tilde{\phi} \tilde{\phi}}$ takes a negative value, the
low-$k$ modes with Eq.~(\ref{Exp:TI}) can be enhanced by the tachyonic
instability. However, the tachyonic instability
is not sustainable, because $\tilde{V}_{\tilde{\phi} \tilde{\phi}}$ flips
the signature when the oscillating homogeneous mode crosses the
inflection point(s). Because of that, $\omega_k^2$ also flips the signature for the low-$k$ modes,
letting the potential ${\cal V}$ flaps during each oscillation of the
homogeneous mode. In particular, every time $\omega_k^2$ changes the
signature, the adiabatic condition is significantly violated, taking a
large value of $|\frac{d \omega_k}{d \tilde{t}}/\omega^2_k|$. The violation of the
adiabatic condition is the crucial difference between the instability in
the phase 2 and the instabilities in the phase 1 and 3.

\begin{figure}[tbh]
\begin{center}
\begin{tabular}{cc}
\hspace{-20pt} \includegraphics[width=9cm]{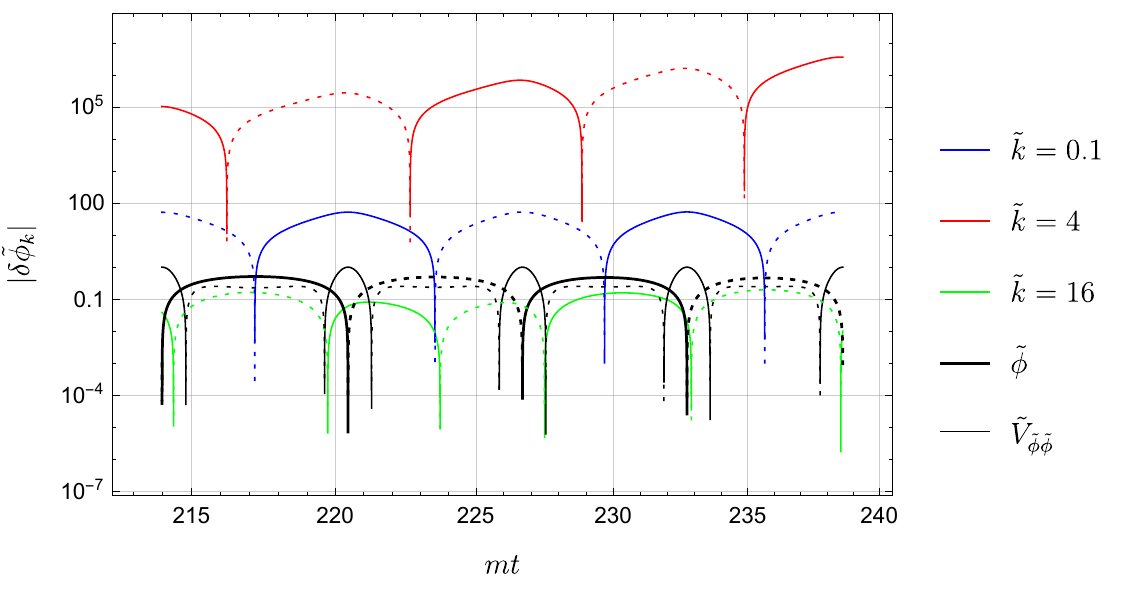}
\includegraphics[width=6.8cm]{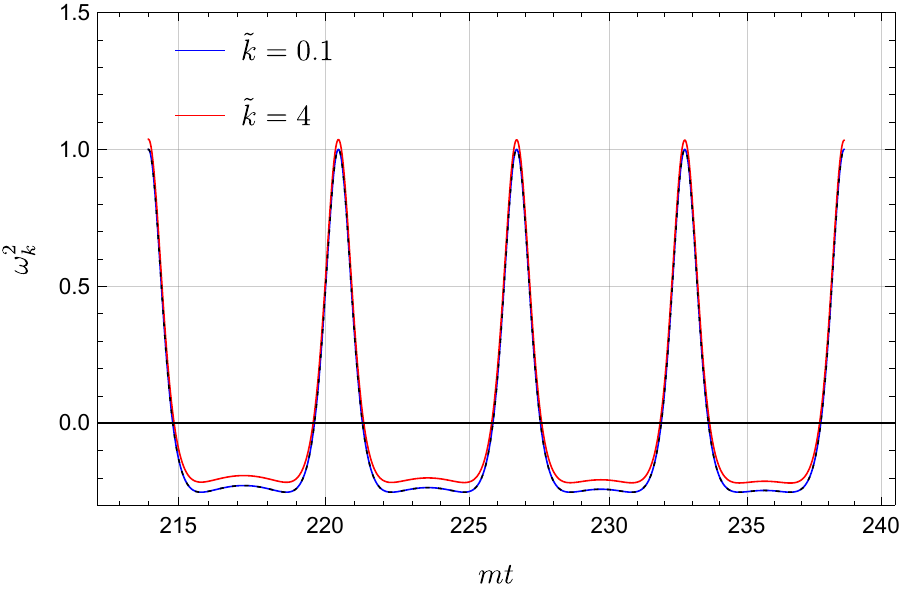}
\end{tabular}
\caption{
The left panel shows the time evolution of $\delta\tilde\phi_k$ for different wave
 numbers $\tilde{k}=k/(a_{H=m}m)=0.1$ (blue), $\tilde{k}=4$ (red), and
 $\tilde{k}=16$ (green) during radiation domination. Here, we consider the $\alpha$ attractor
 potential with $c=5$ under the slow-roll initial condition with $\tilde{\phi}_i = 3$. We also
 showed the time evolution of $\tilde{\phi}$ (black thicker) and
 $\tilde{V}_{\tilde{\phi} \tilde{\phi}}$ (black thinner). We
 distinguished positive values and negative values, using solid lines for
 the former and dotted lines for the latter. The right panel shows the
 time evolution of $\omega^2_k$ for $\tilde{k}=0.1$ (blue) and
 $\tilde{k}=4$ (red). The black dotted line shows
 $\omega_{k=0}^2=\tilde{V}_{\tilde{\phi} \tilde{\phi}}$, which coincides
 with the one for $\tilde{k}=0.1$.}
\label{Fg:focusphic5}
\end{center}
\end{figure}

For the conventional broad resonance, inhomogeneous modes are largely
enhanced at the moment when the adiabatic condition is violated. By
contrast, for the instability in the phase 2, the enhancement takes
place when $\omega_k^2$ takes negative values and finishes, when it
turns to be positive, violating the adiabatic condition. This can be
seen in Fig.\,\ref{Fg:focusphic5}, which shows a typical time
evolution of $\delta \tilde\phi_k$ (left) and $\omega_k^2$ (right)
during this phase obtained by numerical calculation. Here, we considered  
the $\alpha$ attractor type potential (\ref{Exp:Valpha}) with $c=5$,
choosing the initial condition $\tilde{\phi}_i=3$. For a reference, we also plotted the time evolution
of the background mode, $\tilde{\phi}$ and $\tilde{V}_{\tilde{\phi} \tilde{\phi}}$.
The three different modes with $\tilde{k} \equiv k/(a_{H=m} m) =0.1,\, 4, 16$
evolve in a different way. In the following, we set the scale factor at $H=m$,
$a_{H=m}$, to unity. Since $\omega_k^2$ with 
$\tilde{k}=16$ is always positive (during the time period plotted in
Fig.\,\ref{Fg:focusphic5}), this mode simply oscillates without
tachyonic growth. Meanwhile, $\omega^2_k$
becomes negative during each oscillation of $\tilde{\phi}$ for both
$\tilde{k}=0.1$ and $\tilde{k} =4$. However, only the mildly low-$k$
mode, $\tilde{k} =4$, manages to grow after two oscillations of
$\tilde{\phi}$ plotted in Fig.\,\ref{Fg:focusphic5}.

\begin{figure}[tbh]
\begin{center}
\begin{tabular}{cc}
\includegraphics[width=12cm]{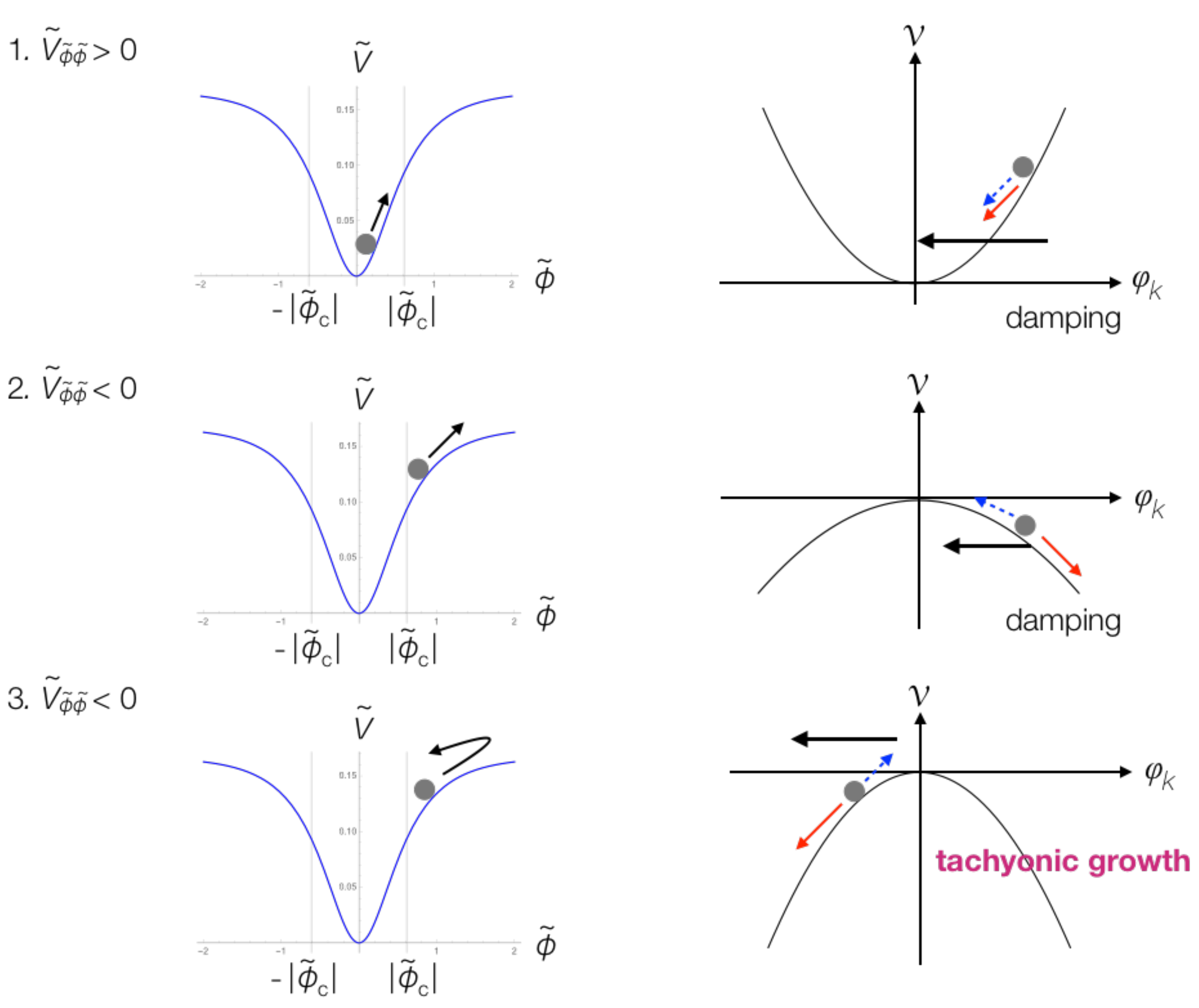}
\end{tabular}
\caption{This figure gives a schematic image of how the low-$k$ modes
 evolve during the phase 2. The blue dotted arrow shows the ``permanent'' restoring force, determined by the first term in (\ref{eq:Vk}), and the red arrow
denotes the potential driven force. The black thicker arrow shows the velocity, i.e., the direction of the motion, of $\varphi_k$. }
\label{Fg:flapping}
\end{center}
\end{figure}
Figure \ref{Fg:flapping} illustrates the reason for this. The blue dotted arrow
denotes the restoring force, determined by the first term in (\ref{eq:Vk}), and the red arrow
denotes the potential driven force, which changes the signature for the
flapping potential. In the first layer of Fig.~\ref{Fg:flapping},
which corresponds to the left-most moment in Fig.\,\ref{Fg:focusphic5},
since the homogeneous mode is within the inflection points, taking
$\tilde{V}_{\tilde{\phi} \tilde{\phi}} > 0$, the potential driven force
also acts as an additional restoring force, which is independent of the
wavenumber and drives $\varphi_k$ towards the origin. As depicted in the second layer in Fig.~\ref{Fg:flapping}, 
once the homogeneous mode passes beyond the inflection point, the flapping potential immediately becomes convex.
During this term, $\varphi_k$ climbs the potential hill, consuming the
kinetic energy obtained in the previous stage. Once $\varphi_k$ crosses
the origin, as depicted in the third layer, $\varphi_k$ goes down the
potential hill, increasing the amplitude through the tachyonic
instability. The tachyonic growth lasts, until the zero mode comes
within the inflection points and the curvature of ${\cal V}_k$, given by
$\omega^2_k$, turns to be positive as in the first layer of
Fig.~\ref{Fg:flapping}. The flow of these steps repeats during the phase
2, where $\tilde{V}_{\tilde{\phi} \tilde{\phi}}$ flips the signature
during each oscillation of the homogeneous mode.

The key to understand the difference between the two modes $\tilde{k}=0.1$
and $\tilde{k}=4$ is in the presence of the second layer stage, i.e., the
amplitude of $\varphi_k$ does not start to grow immediately after
$\tilde{V}_{\tilde{\phi} \tilde{\phi}}$ becomes negative. Reducing the
duration of this stage leads to a net growth, acquiring a longer time
for the tachyonic growth in the third layer stage. Notice that in the
second layer stage, the potential driven force acts as a resistive force
against the motion of $\varphi_k$ towards the origin, while the
``permanent'' restoring force, which is proportional to $(k/am)^2$,
supports the motion towards the origin. Because of that, as shown in
Fig.\,\ref{Fg:focusphic5}, (the amplitude of) the mode $\tilde{k} = 4$
turns to grow prior to $\tilde{k}= 0.1$ after
$\tilde{V}_{\tilde{\phi} \tilde{\phi}}$ becomes negative. For
$\tilde{k}=0.1$, since the growth during the third layer stage is canceled by
the decay in the first layer and the second layer stages, there is no net growth.

We dub the instability in the phase 2, which resonates with the flapping $\omega^2_k$
or ${\cal V}_k$, {\it the flapping resonance instability},
distinguishing it from the usual broad resonance instability and also from
the usual tachyonic instability. Unlike the usual broad
resonance instability and the tachyonic instability, the flapping
resonance generates a peak at $ k \neq 0$ in the spectrum. This is
because the tachyonic instability takes place only for low-$k$
modes and a larger $k$ mode can be enhanced in a longer time among these low-$k$
modes. Therefore, the peak wavenumber generated by the flapping
resonance instability is roughly estimated by the maximum wavenumber
among those which undergo the tachyonic instability as
\begin{align}
 & \frac{k_{\rm peak}}{a_{\rm res} m} \simeq \sqrt{|\tilde{V}_{\tilde{\phi}
 \tilde{\phi}}^{(\rm plat)} |} \,, \label{Exp:peakFR}
\end{align}
where we evaluated $\tilde{V}_{\tilde{\phi} \tilde{\phi}}$ in the
plateau region, which can be reached when $\tilde{\phi}$ climbs up the
potential $\tilde{V}$. Here, $a_{\rm res}$ denotes the scale factor,
when the flapping resonance takes place.

The flapping resonance generically takes place, when $\omega^2_k$
repeatedly changes the signature. Therefore, when the scalar potential
of the inflation satisfies the conditions i) and ii), the flapping
resonance also takes place during reheating (see also Ref.~\cite{Antusch:2015nla}). In the early stage of reheating after small
field inflation models, $\omega^2_k$ flips the signature during each oscillation of the inflaton~\cite{Brax:2010ai}. In this case, unlike the plateau case, the
inflaton starts to oscillate around $H_{\rm osc} \simeq m$. Then, this stage does not last long, since the inflaton exits the tachyonic
region after several oscillations due to the Hubble friction~\cite{Brax:2010ai}.

\subsubsection{Narrow resonance: Phase 3} \label{SSSec:NR}
In this phase, the homogeneous mode of the axion oscillates within the inflection points.
During this phase, we can expand perturbatively the second derivative of the potential as
\begin{equation}
 \tilde{V}_{\tilde{\phi} \tilde{\phi}} =  1+\frac{1}{2} \lambda \tilde\phi^2 +O(\tilde\phi^4) 
\end{equation}
in a good approximation. For the $\alpha$ attractor type
potential (\ref{Exp:Valpha}), $\lambda$ is given by $\lambda = - 4(2 + 3c)$. 
In this phase, the evolution of the homogeneous mode can be well
approximated as $\tilde{\phi} = \tilde{\phi}_* (a_*/a)^{3/2} \cos \tilde{t}$, where
quantities with $*$ are evaluated at the begining of the phase 3.
Using this expression, the mode equation can be rewritten as the Mathieu equation,
\begin{align}
 & \frac{d^2 \varphi_k}{d \tilde{t}^2} + \left[ A_k - 2 q \cos 2\tilde{t} \right]
 \varphi_k = 0 \label{Eq:Mathieu}\,,
\end{align}
where we introduced 
\begin{align}
 & q \equiv  - \frac{\lambda}{8}\, \tilde{\phi}_*^2 \left( \frac{a_*}{a} \right)^3 \,,  \label{Exp:qz} \\[2mm]
 & A_k \equiv 1 + \left( \frac{k}{am} \right)^2 - 2 q\,.
\end{align}
The analysis based on the Mathieu equation is possible only in the phase
3, where the homogeneous mode can be well approximated by the harmonic
oscillation.

The solution of the Mathieu equation has resonance bands around 
$A_k \simeq n^2$ with $n= 1,\, 2,\, \cdots$. In particular,
the first resonance band is given by $A_k - 1 = \pm q$. This implies that for
$\lambda < 0$, the resonance instability takes place for the physical
wavenumbers which satisfy 
\begin{align}
 & q < \left( \frac{k}{am} \right)^2 < 3 q\,. \label{Eq:RB} 
\end{align}
At $a \simeq a_*$, one obtains the peak wavenumber in the first
band as
\begin{align}
 & \frac{k_{\rm peak}}{a_* m} \simeq \sqrt{2q} \simeq
\frac{\sqrt{|\lambda|} |\tilde{\phi}_*|}{2}\,.  \label{Exp:peakNR}
\end{align}
In this phase, $\omega^2_k$ stays all the time positive and
there is no violation of the adiabatic condition, which indicates that
the instability in this phase is the narrow resonance instability. The
narrow resonance instability in the phase 3 proceeds much more slowly
than the flapping resonance instability in the phase 2.

When the oscillation starts around $H_{\rm osc} \simeq m$, the
wavenumbers in the instability bands are red-shifted away after several
oscillations. By contrast, in case $H_{\rm osc} \ll m$, the scale factor
almost does not change in one oscillation period of the zero mode and
the instability modes stay in the resonance bands over many periods
of oscillation. In addition, since the amplitude of
$\tilde{\phi}$ and $q$ do not decrease over many periods of the
oscillation for $H_{\rm osc}/m \ll 1$, the growth rate and the width of the
resonance band do not decrease either in the time scale of the oscillation.

During the phase 2, the narrow resonance instability is not very efficient
and it can be important only after entering the phase 3. This is because
the oscillation period of the homogeneous mode $\tilde{\phi}$,
determined by $\tilde{V}_{\tilde{\phi} \tilde{\phi}}$, still
considerably changes in time during the phase 2. Since the Bose enhancement is caused by
acting a regulated periodic force over many oscillation periods, the
parametric resonance instability is not very efficient during the phase 2.

\subsection{Overall evolution in linear regime}
\begin{figure}[tbh]
\begin{center}
\begin{tabular}{cc}
\hspace{-20pt}\includegraphics[width=9.2cm]{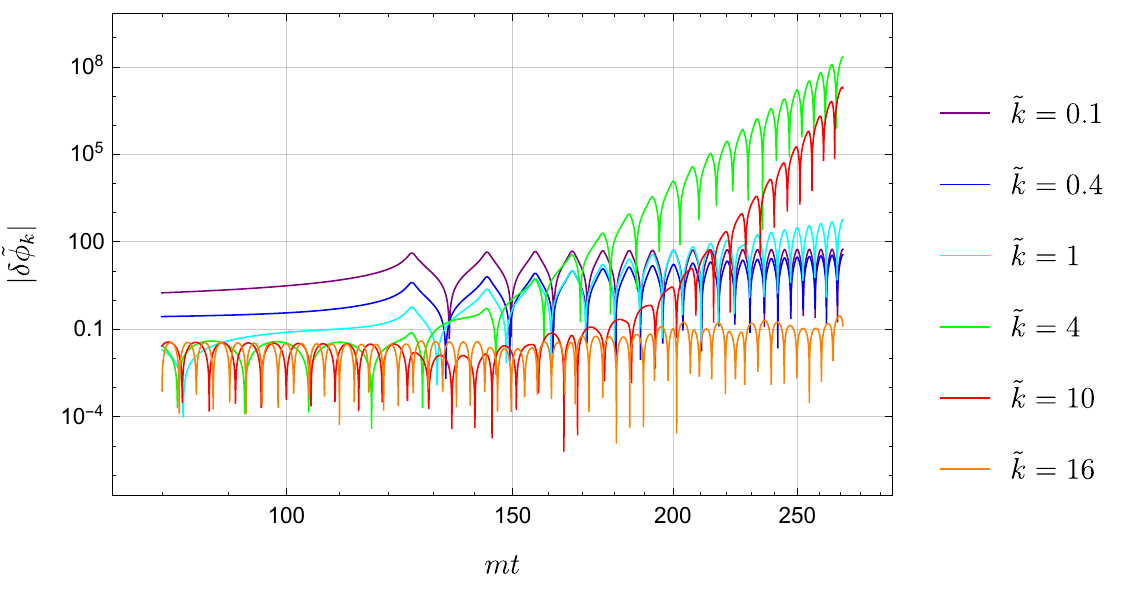}
\hspace{-5pt}\includegraphics[width=7cm]{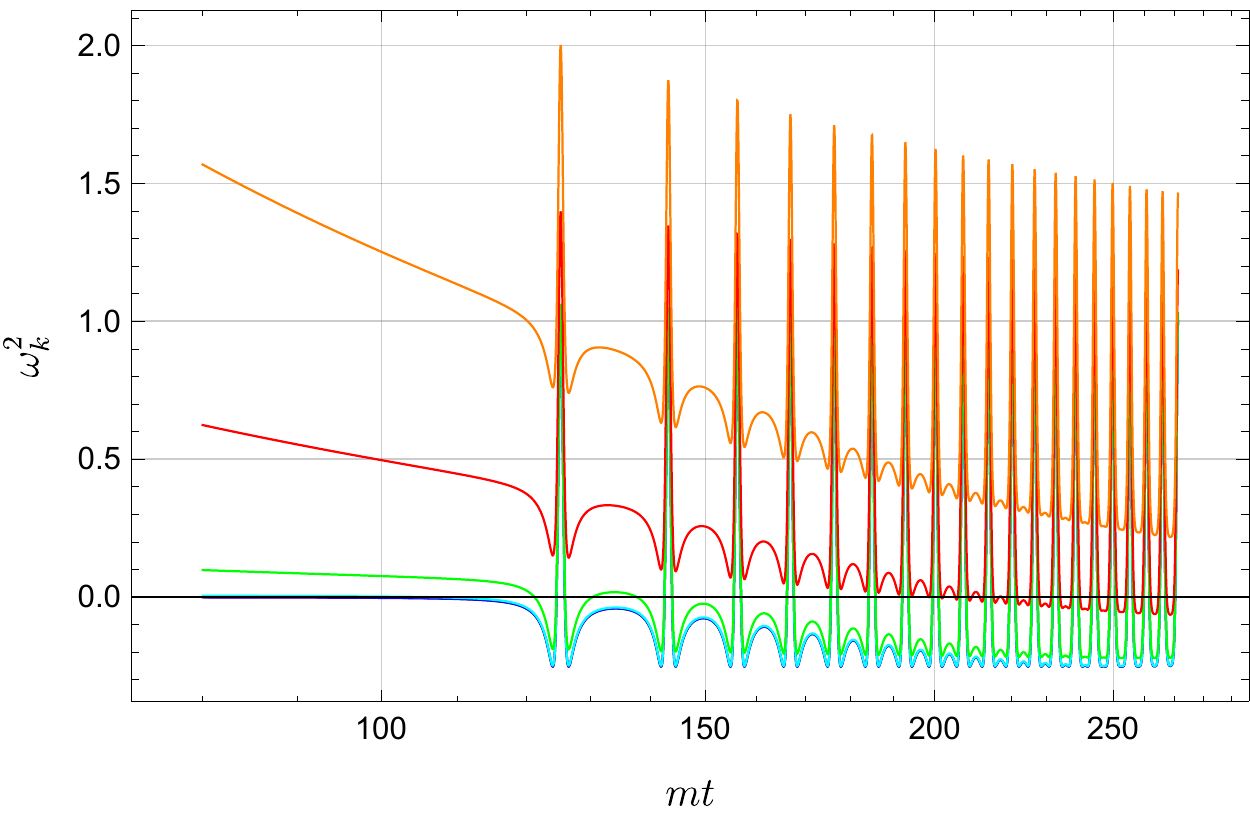}
\end{tabular}
\caption{
The left panel shows the time evolution of $\delta\tilde\phi_k$ for the
 $\alpha$ attractor potential with $c=5$ and the initial condition
 $\tilde{\phi}_i=3$. The right panel shows the time evolution of
 $\omega^2_k$ during the same time period.
}
\label{Fg:fgc5p3}
\end{center}
\end{figure}
\begin{figure}[tbh]
\begin{center}
\begin{tabular}{cc}
\hspace{-20pt}\includegraphics[width=9.2cm]{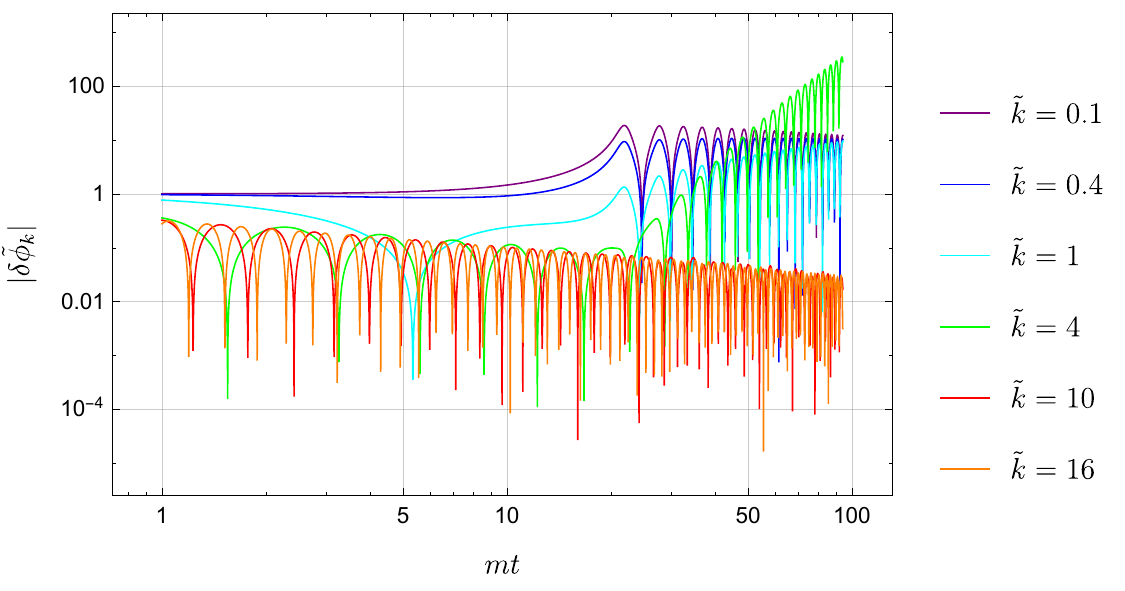}
\hspace{-5pt} \includegraphics[width=7cm]{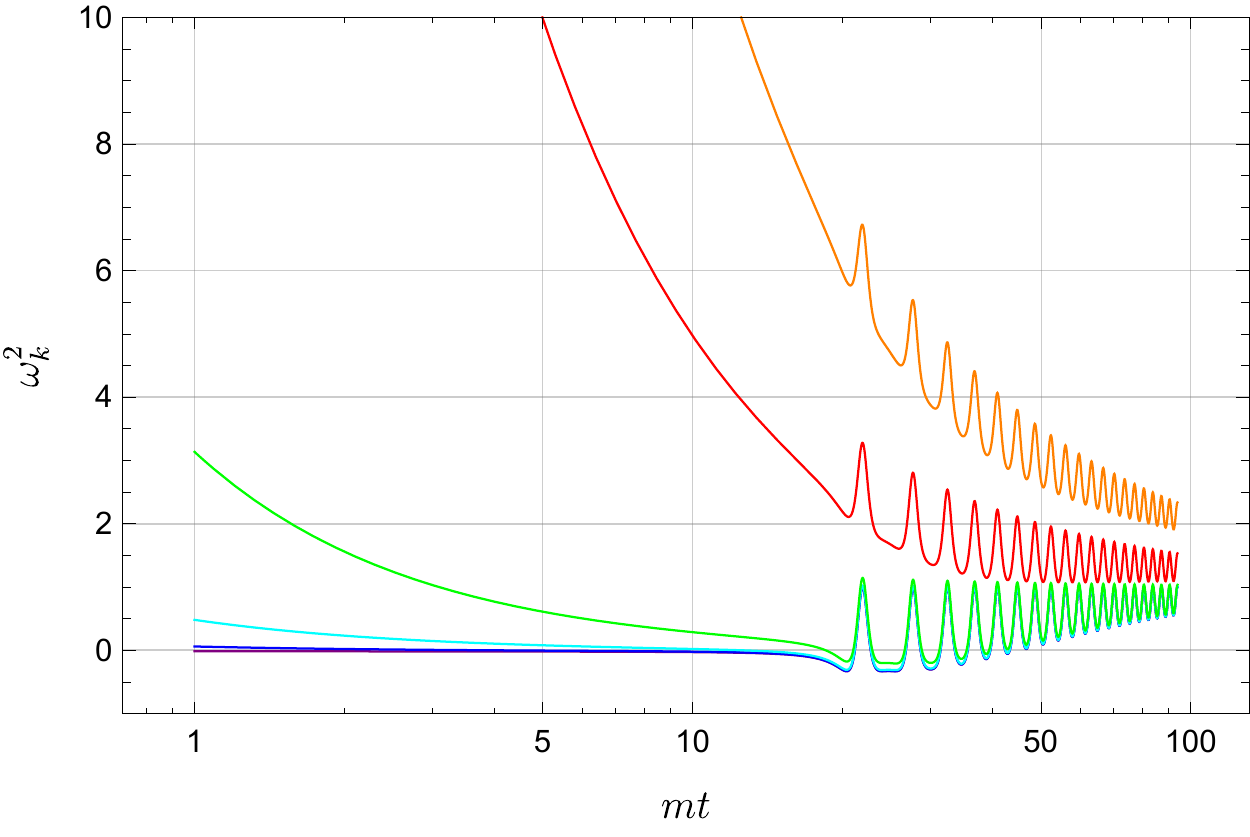}
\end{tabular}
\caption{
Same as Fig.~\ref{Fg:fgc5p3} except that $c$ is now set to $c=0$.
}
\label{Fg:fgc0p3}
\end{center}
\end{figure}
In this section, we showed that the different types of instabilities
become prominent during the different phases. The left panels of Fig.~\ref{Fg:fgc5p3} and
Fig.~\ref{Fg:fgc0p3} show two typical time evolutions of 
$\delta \tilde{\phi}_k$ for potentials which satisfy the two conditions
i) and ii). Here, again as an example, we considered the
$\alpha$-attractor potential with $c=5$ (Fig.~\ref{Fg:fgc5p3})
and $c=0$ (Fig.~\ref{Fg:fgc0p3}). In both cases, we chose
the initial condition as $\tilde{\phi}_i=3$ under the slow-roll
approximation, which is the attractor solution in the plateau region.   
The right panels show the time evolution of $\omega_k^2$ for each
wavenumber.

When the homogeneous mode $\tilde{\phi}$ rolls down the plateau region (phase 1), $\omega^2_k$ for
the low-$k$ modes such as $\tilde{k} = 0.1$ and $\tilde{k} = 0.4$ stay
negative, leading to the tachyonic instability. Once $\tilde{\phi}$
starts to oscillate, $\omega_k^2$ flips the signature, leading to the
flapping resonance instability. The phase 2 for $c=5$ continues longer
than the one for $c=2$ as is shown in the right panels of
Fig.~\ref{Fg:fgc5p3} and Fig.~\ref{Fg:fgc0p3} (this can be clearly seen
e.g., by looking at $\tilde{k}= 0.1$, for which
$\omega_k^2 \simeq \tilde{V}_{\tilde{\phi} \tilde{\phi}}$). One of the reasons for this is because for $c=5$ the onset
of the oscillation is delayed more and the Hubble friction is less efficient. It is, however, not entirely clear what controls the duration of the phase 2 and this will be investigated in our future study.  During the phase 3, the narrow resonance sets in. The growth rate of the narrow resonance is smaller than the flapping resonance. 
When the phase 2 continues sufficiently long, the energy
transfer from the homogeneous mode to the inhomogeneous mode finishes
before entering the phase 3. Therefore, depending on how long the phase
2 persists, the dominant instability and the resultant spectrum will be
different (see Table \ref{Table}).

\begin{table}[t]
	\centering
	\begin{tabular}{|c|c|c|}
		\hline	
        & No delay ($ H_{\rm osc} /m \simeq 1$) & Delay ($ H_{\rm osc} /m \ll 1$) \\
\hline 
 Phase 2, short  & No instability & Tachyonic ($\to$ Flapping res.) $\to$ Narrow res.  \\
		\hline 
Phase 2, long &  -  & Tachyonic $\to$ Flapping res.  \\
		\hline 
	\end{tabular} 
	\caption{This table summarizes the different instability
 processes for different setups. By the terms ``long'' and ``short'',
 we mean the phase 2 continues over many periods of the oscillation or
 finishes after several oscillations.}\label{Table}
\end{table}

\section{Lattice simulation}

In the previous section, we discussed the evolution of the inhomogeneous mode
based on linear analysis. When we neglect the non-linear contributions, the energy
transfer from the homogeneous mode to the inhomogeneous modes eternally
continues, which is obviously wrong. In fact, once the inhomogeneous
modes become comparable to the homogeneous mode, the backreaction and
the rescattering turn on and the dynamics enters a highly non-linear
regime. In this section, we address the non-linear dynamics based on the lattice simulation.

\subsection{Nonlinear dynamics of the axion}
Here we solve the field equation (\ref{KG}) directly in the lattice space where the spatial derivative is replaced with the finite difference.
For computational convenience, we use the conformal time, $\tau$ as a time variable, defined by $d\tau = dt/a$ instead of the cosmic time and in order to remove the Hubble friction term, we redefine the field variable as $\tilde\phi = \tilde\Phi/a$. Then, the equation of motion (\ref{KG}) can be rewritten as
\begin{equation} \label{eq:eom_Phi}
\tilde\Phi''-\partial_{\tilde{\sbm{x}}}^2 \tilde\Phi -\frac{a''}{a} \tilde\Phi +a^3 \tilde{V}_{\tilde\phi} = 0,
\end{equation}
where the prime ($\tilde\Phi'$) expresses the derivative with respect to the normalized conformal time $m\tau$.
We have solved the above equation by using fourth order Symplectic
integrator with $256^3$ grids for the $\alpha$ attractor type potential
in the radiation dominated Universe, $a\propto \tau$. We start to solve
the lattice simulation at the time when $H=m$, again setting 
$a_{H=m} = 1$, i.e. $m\tau_i=1$, and imposing the slow-roll condition\footnote{As was
discussed in the previous section, this can be verified in the case of our interest where the
axion was initially located at the plateau region. }. The simulation box size is set to $2\pi m^{-1}$.

\begin{figure}[tp]
\centering
\subfigure[$c=5,\tilde\phi_i=3$]{
\includegraphics [width = 7cm, clip]{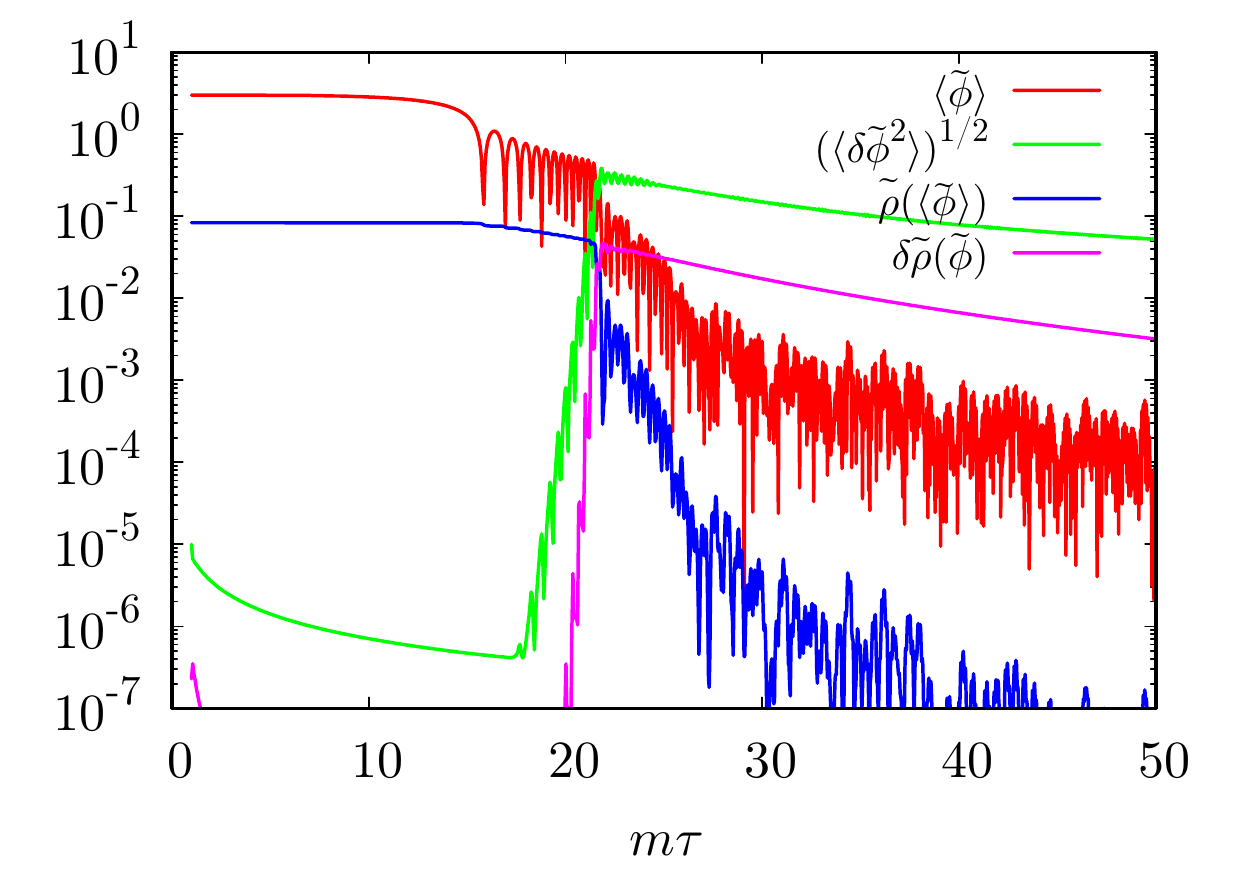}
\label{subfig:evolve1}
}
\subfigure[$c=2,\tilde\phi_i=3$]{
\includegraphics [width = 7cm, clip]{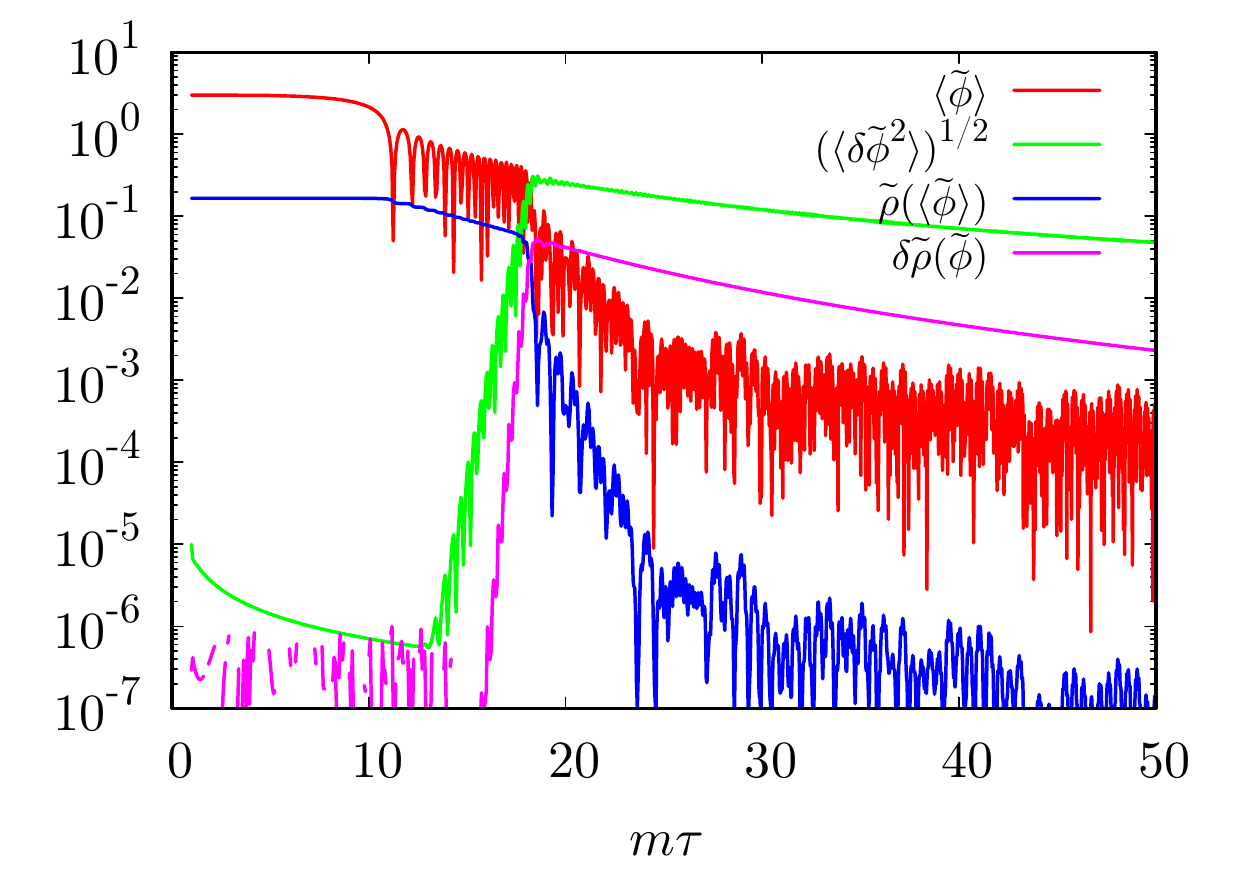}
\label{subfig:evolve2}
}

\caption{
Evolution of the spatial average of $\tilde\phi$ (red), the root-mean-square $\langle \delta\tilde\phi^2 \rangle^{1/2}$ (green), the energy density of the average $\tilde\rho(\langle \tilde\phi \rangle)$ (blue), energy density perturbation, $\langle \delta\tilde\rho^2 \rangle^{1/2}$ (orange). We have taken $c=5$ ($c=2)$ and $\tilde\phi_i=3$ in the left (right) panel.
}
\label{fig:evolve12}
\end{figure}

\begin{figure}[tp]
\centering
\subfigure[$c=0,\tilde\phi_i=3$]{
\includegraphics [width = 7cm, clip]{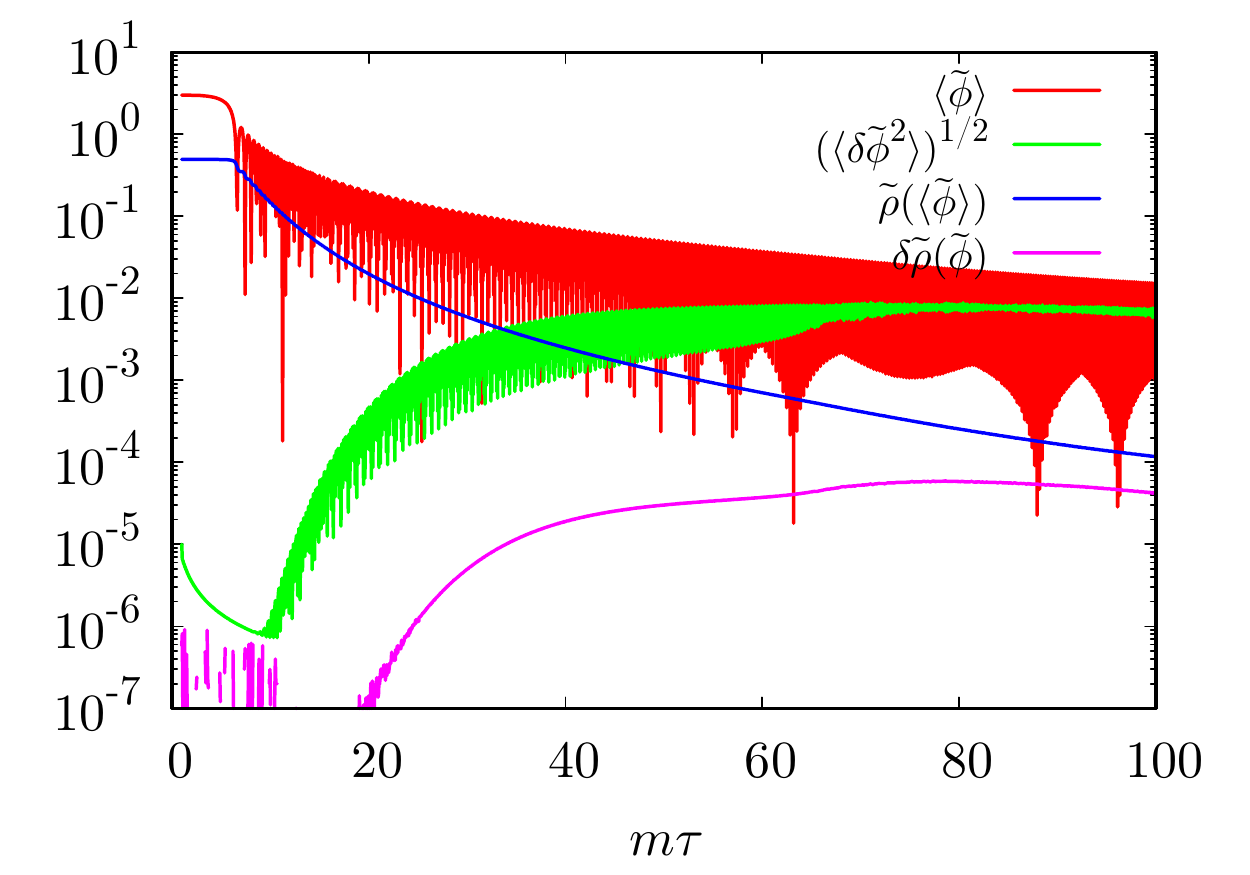}
\label{subfig:evolve3}
}
\subfigure[$c=2,\tilde\phi_i=2$]{
\includegraphics [width = 7cm, clip]{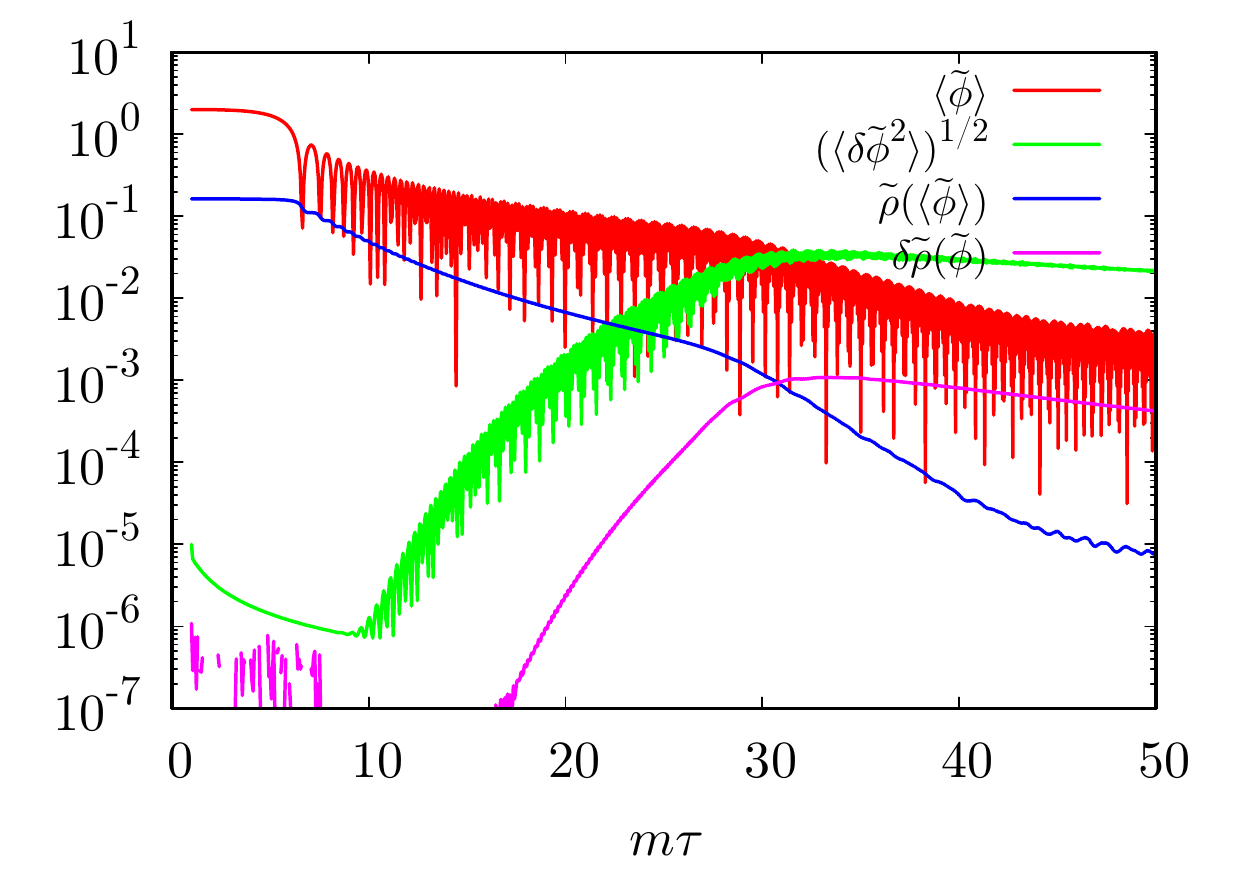}
\label{subfig:evolve4}
}

\caption{
Same as Fig.\,\ref{fig:evolve12} but $c=0$ and $\tilde\phi_i=3$ in the left panel and $c=2$ and $\tilde\phi_i=2$ in the right panel.
}
\label{fig:evolve34}
\end{figure}

Fig.\,\ref{fig:evolve12} and \ref{fig:evolve34} show the evolution of
the homogeneous mode, i.e. spatially-averaged field value (red), the
root-mean-squared of the field fluctuation (green), the energy density
of the homogeneous mode (blue) and the energy density fluctuation
(magenta). After the (delayed) onset of the axion oscillation, the field
fluctuation grows exponentially and it eventually dominates over the
homogeneous component (except for the case with $c=0$ and
$\tilde\phi_i=3$). At that time, the exponential growth stops due to the backreaction from produced inhomogeneous modes on the homogeneous mode. 
Fig.\,\ref{fig:evolve12} shows that, through the flapping resonance in
phase 2, the fluctuation grows quickly and saturates in a short time
period. On the other hand, Fig.\,\ref{fig:evolve34} shows the growth due
to the narrow resonance regime in phase 3. In this case, the growth rate
is smaller than that of the flapping resonance and it takes longer time
for fluctuations to catch up with the zero mode or fluctuations never catch up with the zero mode as can be seen in Fig.\,\ref{subfig:evolve3}

To see the spectrum of the enhanced nonzero $k$-mode axion, let us consider the Fourier transformation, $\Phi_{\bf k}$, and 
define the occupation number (normalized by $mf^2$) as
\begin{equation}
n_k = \frac{1}{2} \left( \frac{|\tilde\Phi_{\bf k}'|^2}{\Omega_k} + \Omega_k |\tilde\Phi_{\bf k}|^2 \right),
\end{equation}
where $\Omega_k$ is defined by\footnote{In order to avoid imaginary
numbers in our numerical computation, we set 
$\langle \tilde{V}_{\tilde\phi\tilde\phi} \rangle = 1$, which corresponds to the value
when the homogeneous mode of the axion is around the potential minimum during each
oscillation.}
\begin{equation}
\Omega^2_k = \left(\frac{k}{m}\right)^2 + a^2 \langle \tilde{V}_{\tilde\phi\tilde\phi} \rangle.
\end{equation}

\begin{figure}[tp]
\centering
\subfigure[$m\tau=15$--25]{
\includegraphics [width = 7cm, clip]{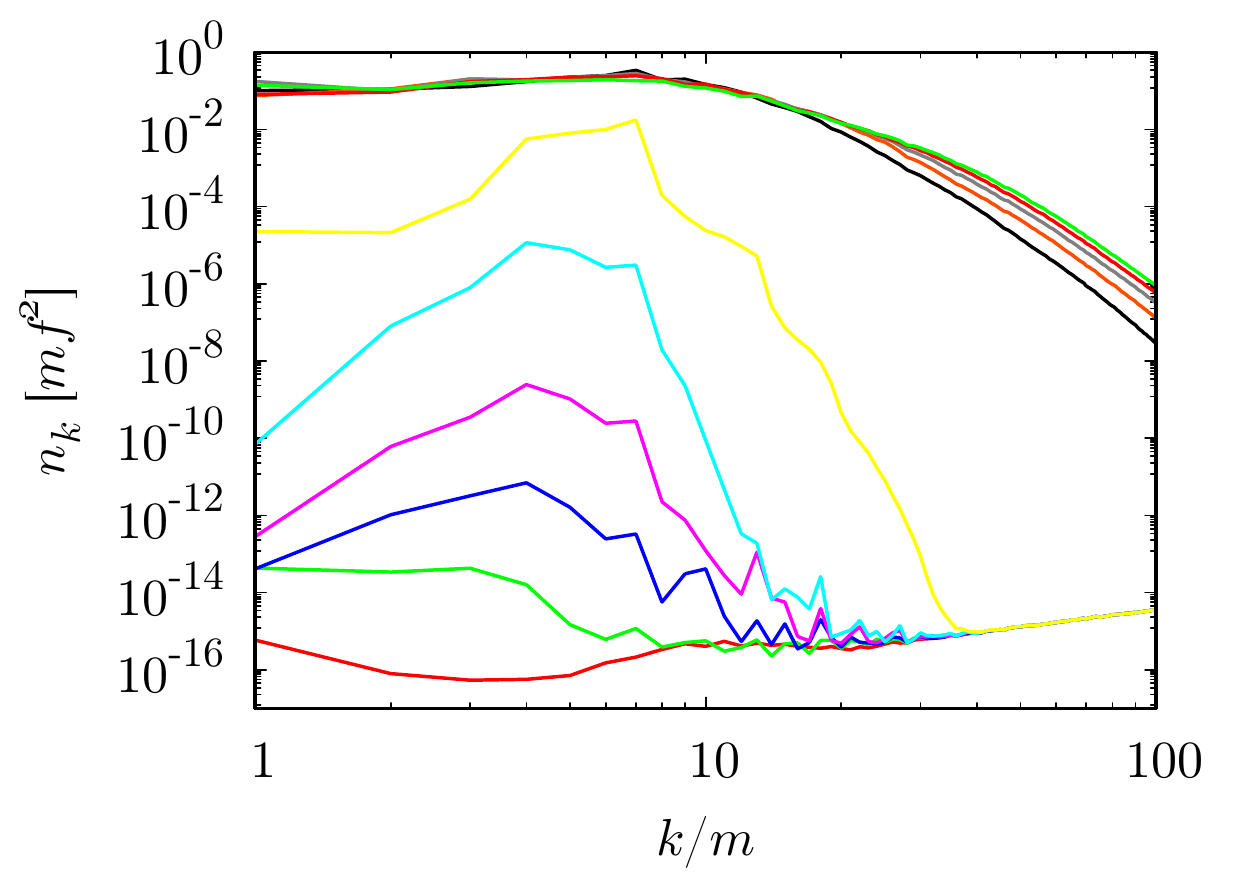}
\label{subfig:nk1a}
}
\subfigure[$m\tau=25$--75]{
\includegraphics [width = 7cm, clip]{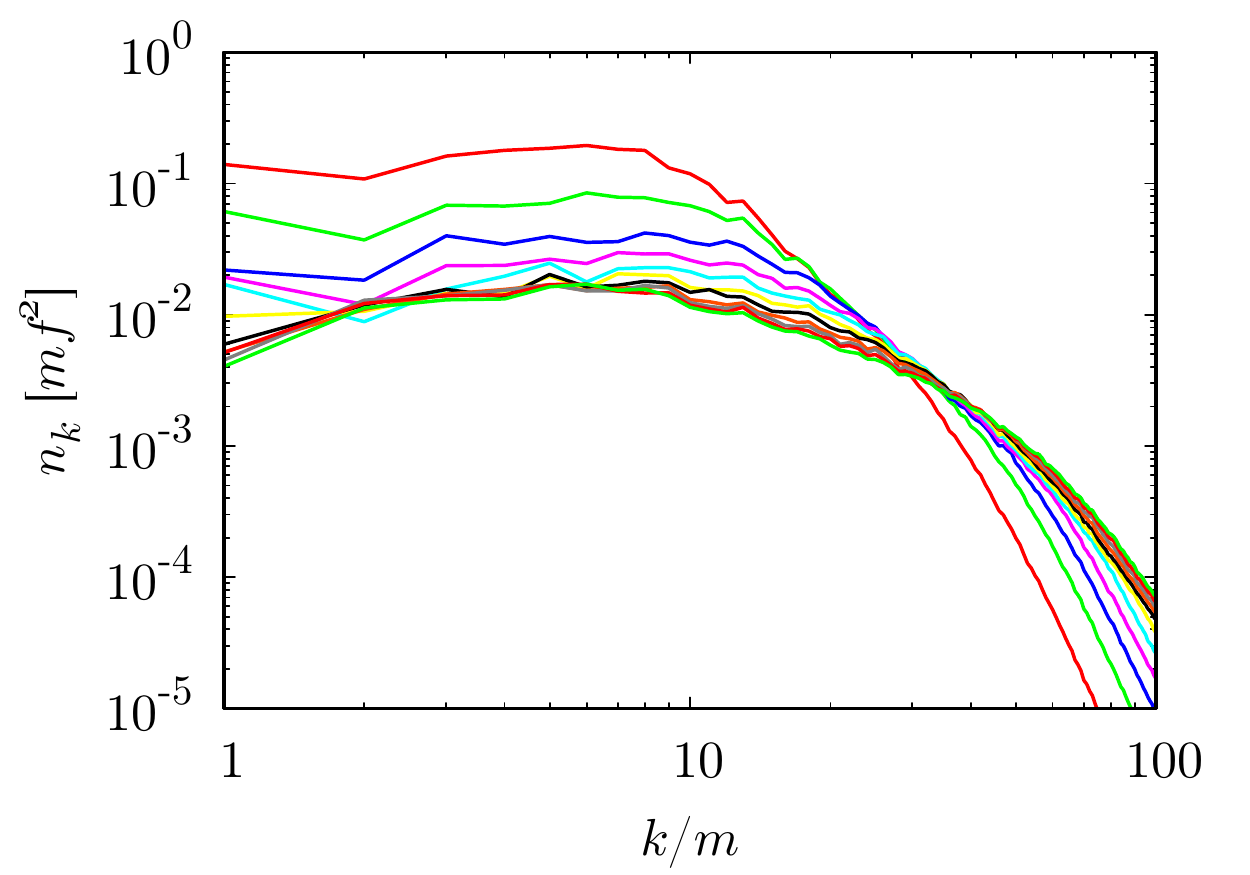}
\label{subfig:nk1b}
}
\caption{
Evolution of the occupation number of the axion for $m\tau = 15$--25 (left), 25--75 (right) in terms of the conformal time. Time evolves from bottom (top) to top (bottom) in the left (right) panel.
We have taken $c=5$ and $\tilde\phi_i=3$.}
\label{fig:nk1}
\end{figure}

\begin{figure}[tp]
\centering
\subfigure[$m\tau=10$--20]{
\includegraphics [width = 7cm, clip]{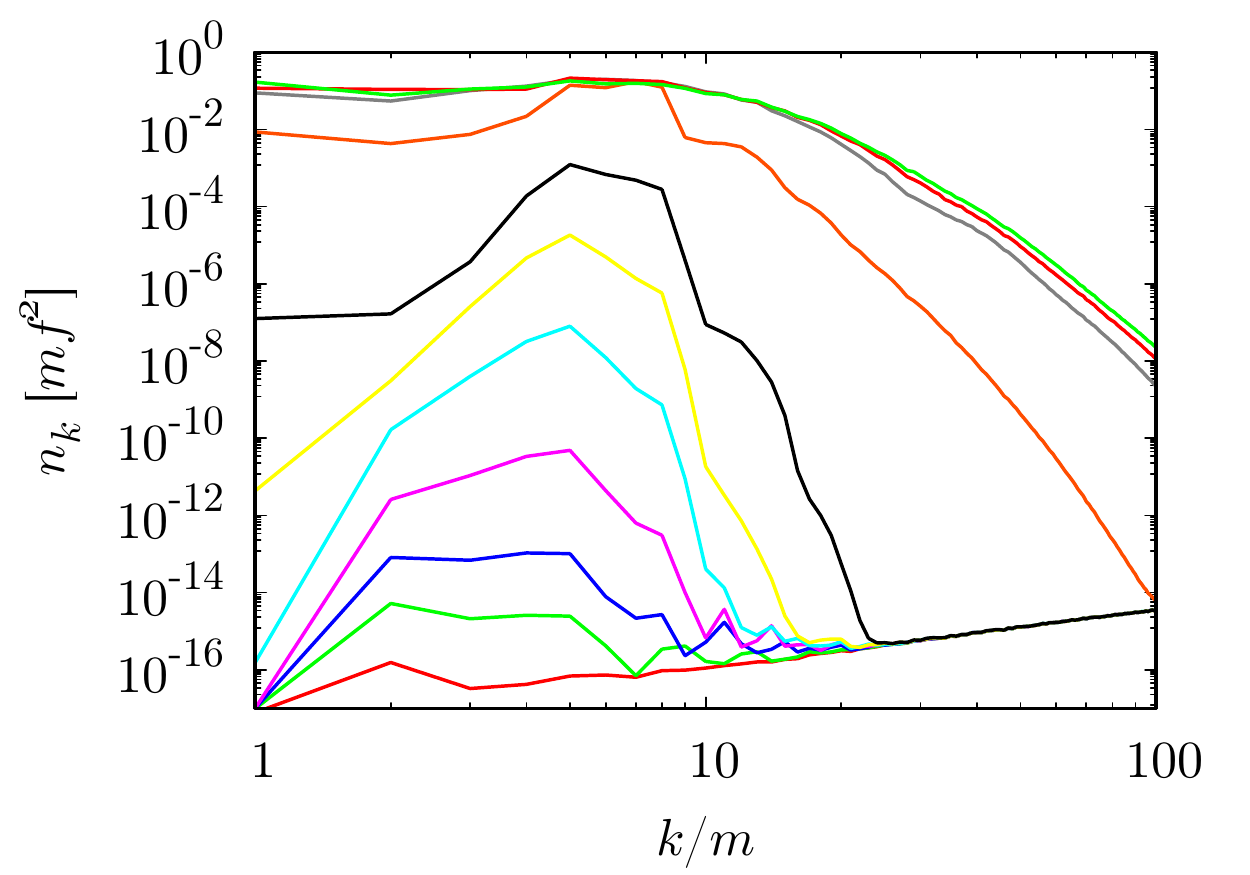}
\label{subfig:nk2a}
}
\subfigure[$m\tau=20$--50]{
\includegraphics [width = 7cm, clip]{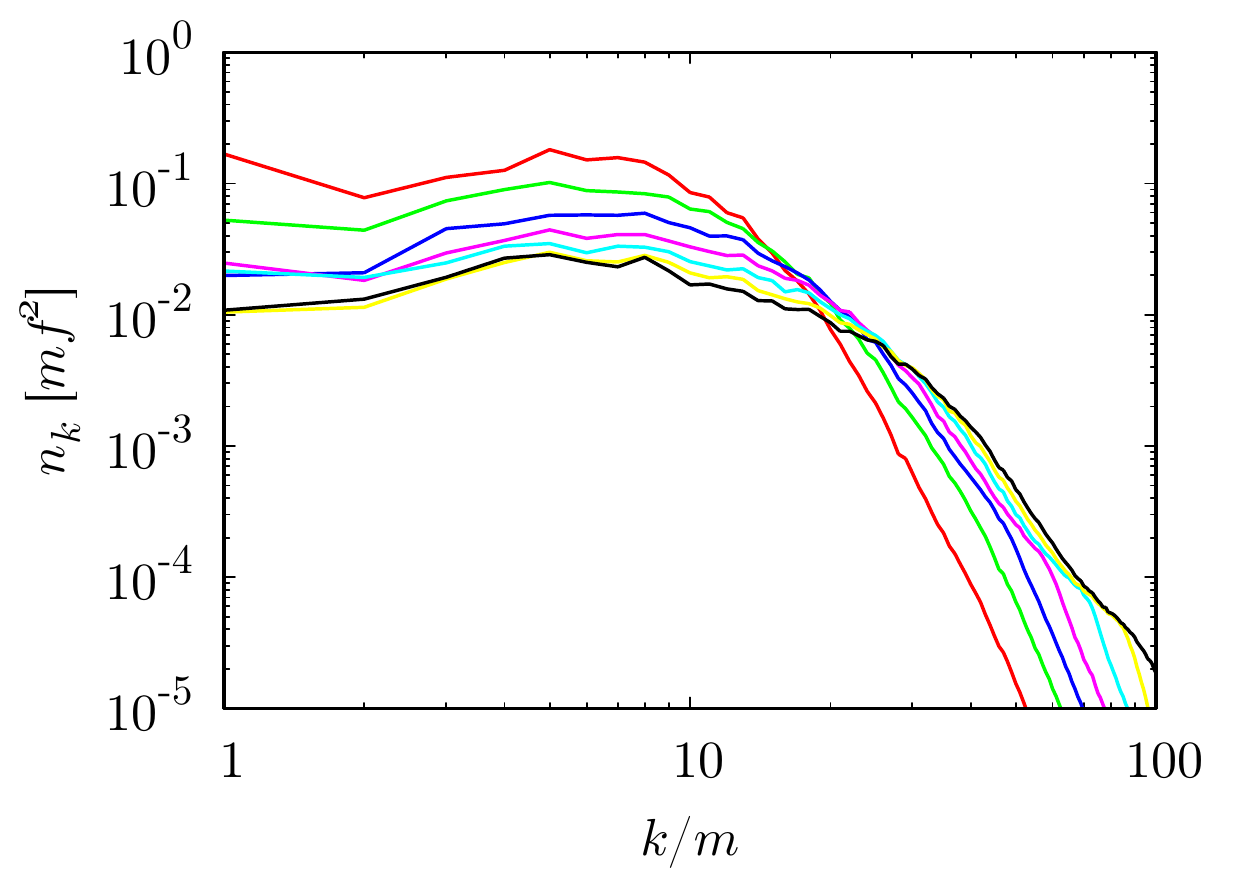}
\label{subfig:nk2b}
}
\caption{
Evolution of the occupation number of the axion for $m\tau = 10$--20 (left), 20--50 (right). Time evolves from bottom (top) to top (bottom) in the left (right) panel.
We have taken $c=2$ and $\tilde\phi_i=3$. }
\label{fig:nk2}
\end{figure}

\begin{figure}[tp]
\centering
\subfigure[$m\tau=10$--65]{
\includegraphics [width = 7cm, clip]{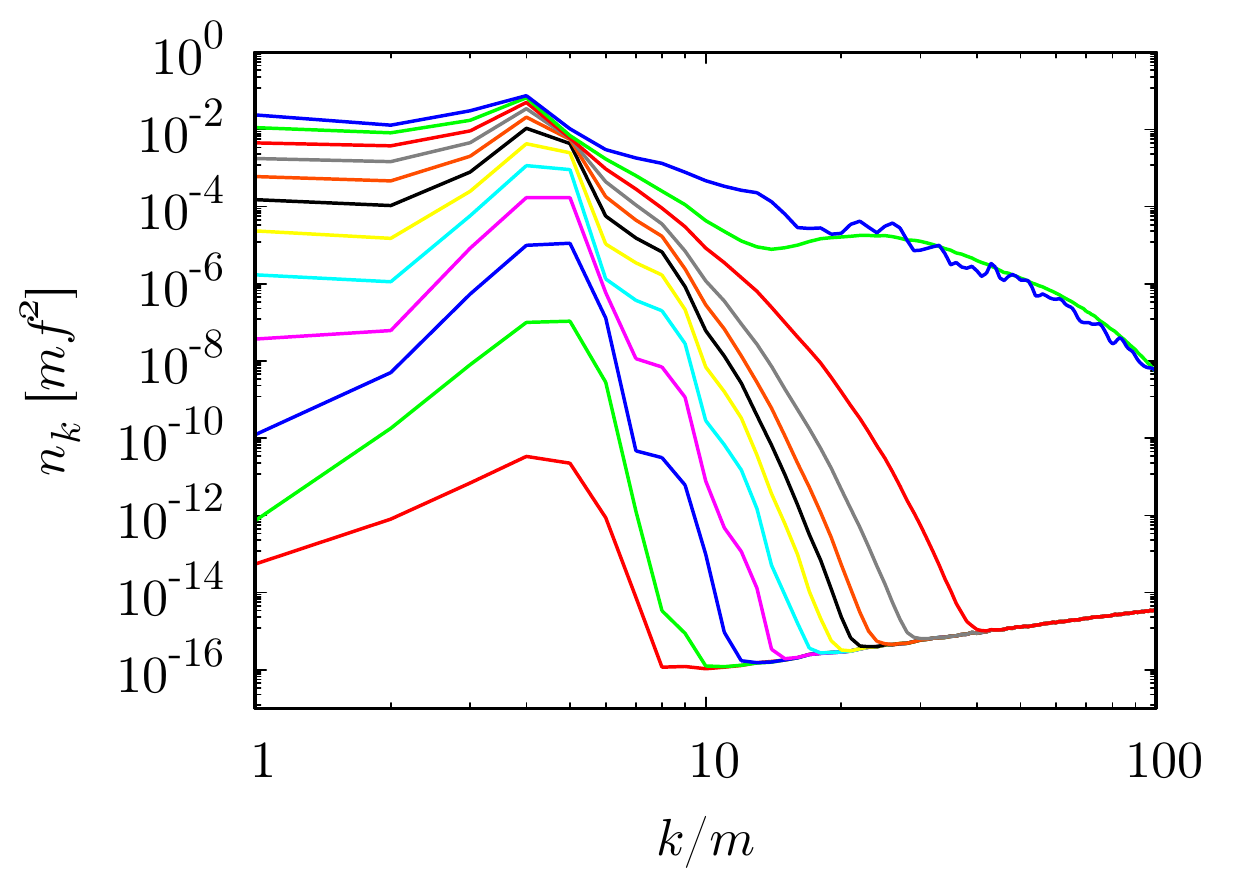}
\label{subfig:nk3a}
}
\subfigure[$m\tau=65$--100]{
\includegraphics [width = 7cm, clip]{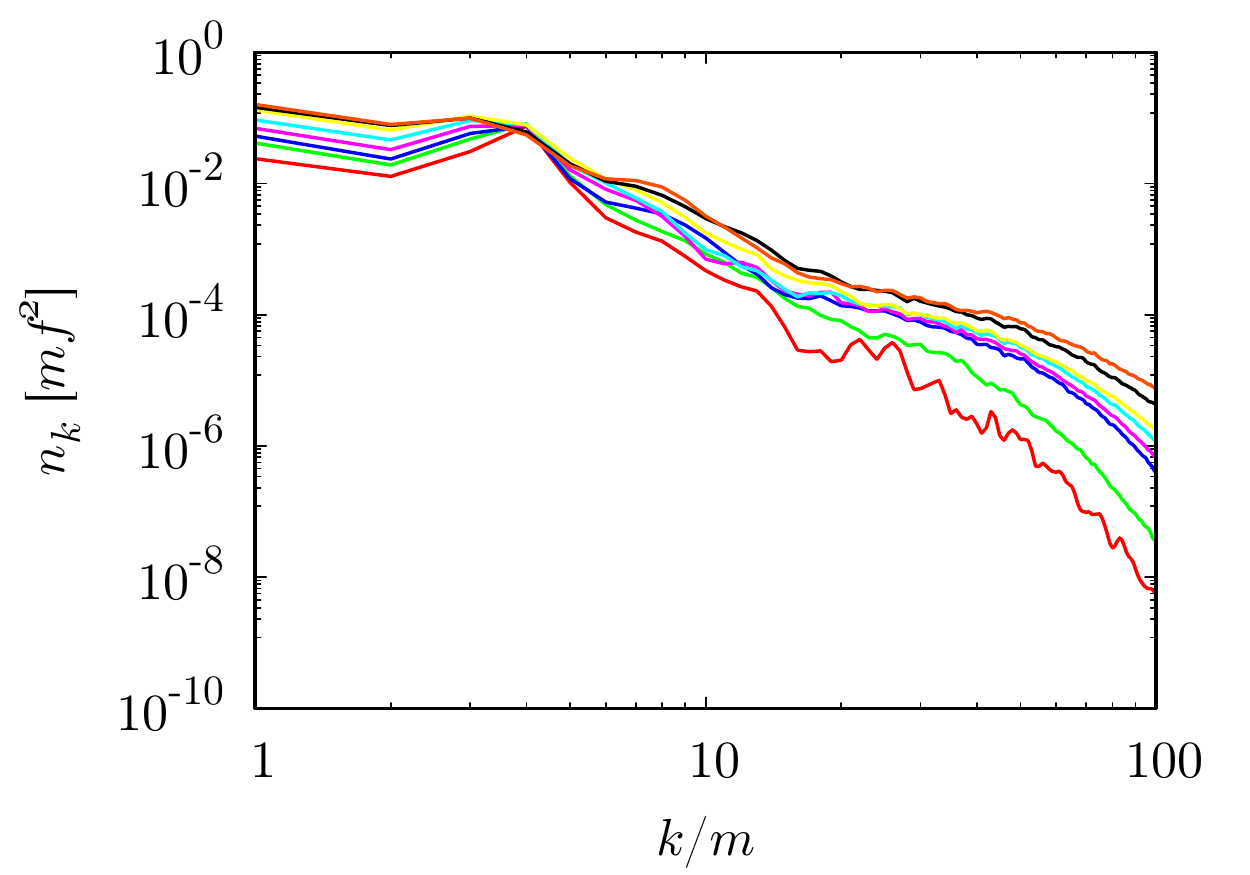}
\label{subfig:nk3b}
}
\caption{
Evolution of the occupation number of the axion for $m\tau = 10$--65 (left), 65--100 (right). Time evolves from bottom (top) to top (bottom) in the both panels.
We have taken $c=0$ and $\tilde\phi_i=3$.}
\label{fig:nk3}
\end{figure}

Fig.\,\ref{fig:nk1}, \ref{fig:nk2}, and \ref{fig:nk3} show the spectrum
of $n_k$. Different curves represent spectra at different time steps. In
Fig.\,\ref{fig:nk1} and \ref{fig:nk2}, the peak modes grow through the
flapping resonance and in Fig.~\ref{fig:nk3}, the peak mode grows
through the narrow resonance. When the energy density in inhomogeneous
modes roughly catches up with the one in the homogeneous mode, the
growth of the inhomogeneous modes terminates and the peaked spectrum
starts to be redistributed by rescattering. As discussed in the context
of reheating, e.g., by Micha and Tkachev~\cite{MT02,MT04}, the turbulence drives the
momentum flow to larger wavenumbers. The flow to UV modes is also found
in our lattice simulation.

\subsection{Oscillon formation}

\begin{figure}[tp]
\centering
\subfigure[$m\tau=18$]{
\includegraphics [width = 7cm, clip]{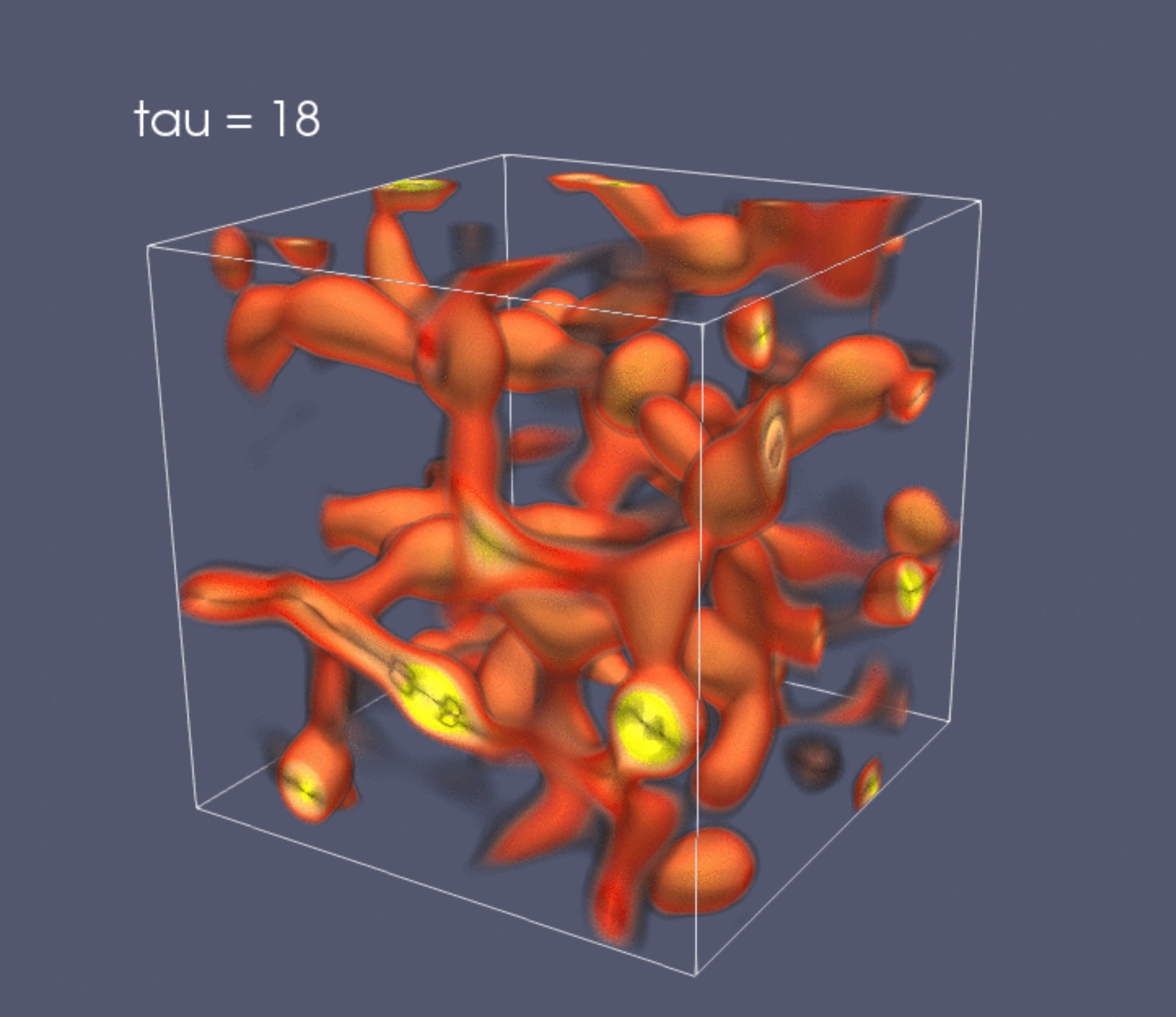}
\label{subfig:o1}
}
\subfigure[$m\tau=20$]{
\includegraphics [width = 7cm, clip]{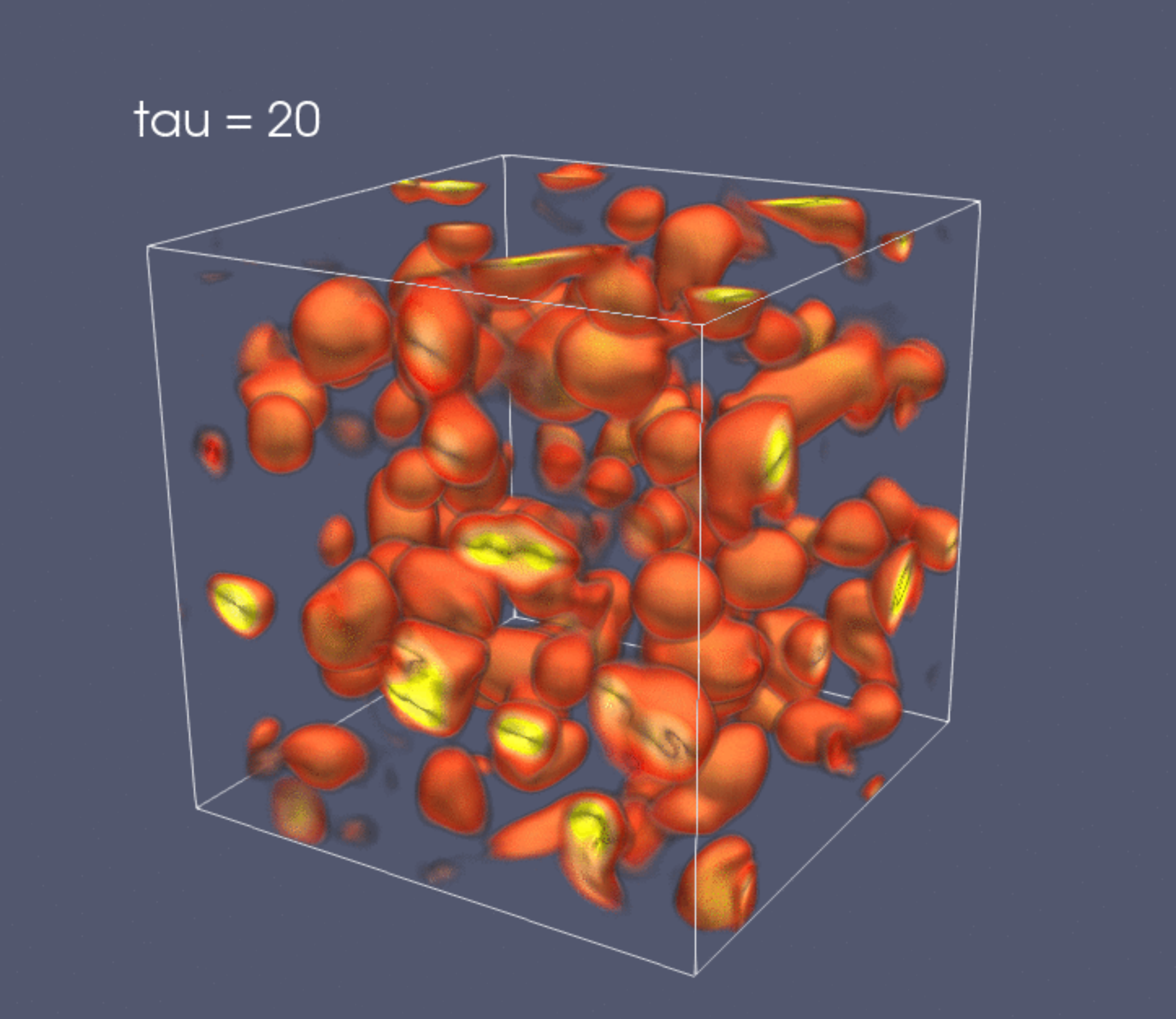}
\label{subfig:o2}
}
\subfigure[$m\tau=40$]{
\includegraphics [width = 7cm, clip]{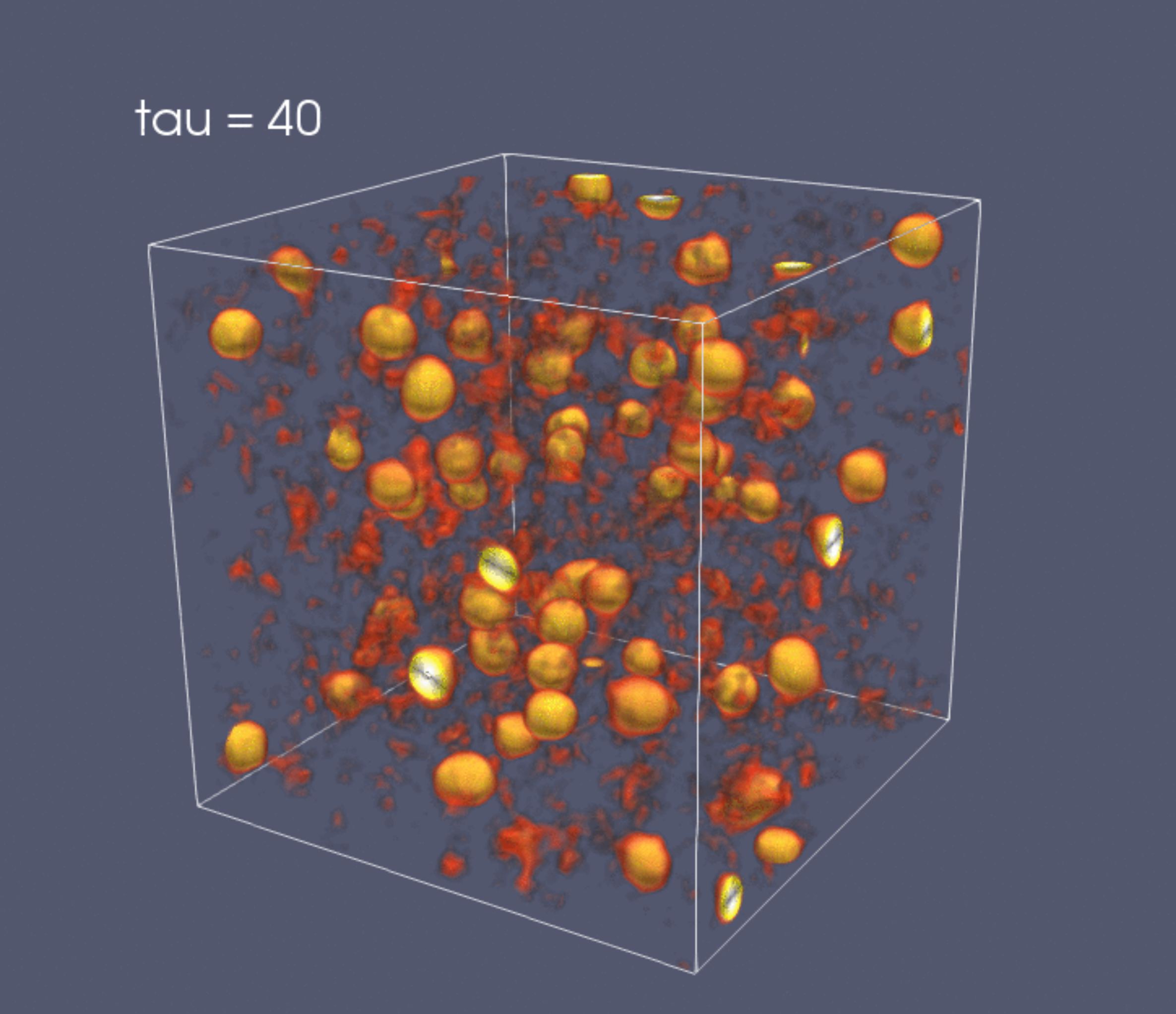}
\label{subfig:o3}
}
\subfigure[$m\tau=50$]{
\includegraphics [width = 7cm, clip]{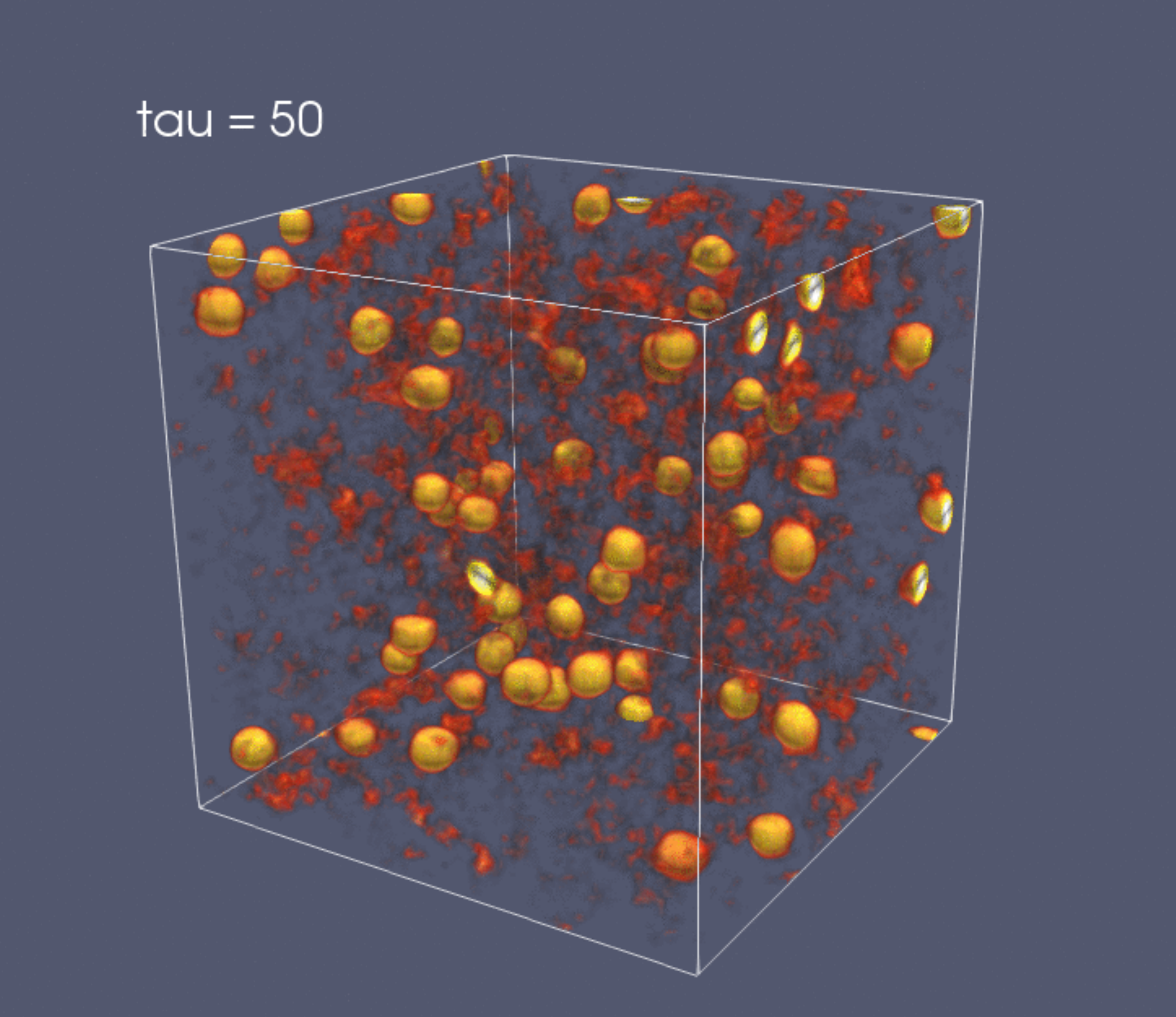}
\label{subfig:o4}
}
\caption{
Snapshots of the evolution of the axion energy density in 3-dimensional lattice space for $m\tau=18$ (upper left), 20 (upper right), 40 (lower left), and 50 (lower right). The red, yellow and white region correspond to $\rho/\bar{\rho} > 2$, $4$ and $10$ respectively with $\bar{\rho}$ being the spatial average of the axion energy density. We have taken $c=5$ and $\tilde\phi_i = 2$.
}
\label{fig:oscillon}
\end{figure}

The plateau condition, the condition ii), which is crucial for a delayed
onset of the oscillation and succeeding instabilities, requires that the absolute value of the
potential gradient, $|\tilde{V}_{\tilde{\phi}}|$, should be smaller than
the one for the quadratic potential, given by $|\tilde{\phi}|$, i.e., the
potential $\tilde{V}$ should be shallower than the one for the quadratic
one. This condition agrees with the often-said condition for the
oscillon formation from various case studies~\cite{oscillon,Amin17}. Indeed, we
found from our lattice simulation that, once the contribution 
from nonzero modes of the axion field becomes sufficiently larger than
that from the homogeneous mode, the axion field forms oscillons, which
are almost spherically symmetric.  
Fig.\,\ref{fig:oscillon} shows snapshots of energy density fluctuation
of the axion. It shows that the oscillon formation occurs around
$m\tau \sim 40$. Note that the energy of the axion is mostly
stored in oscillons after copious oscillons are formed.

\section{GW emission}
In this section, we consider GW emission sourced by axion field
fluctuations. In general, a scalar perturbation does not source GWs in
the linear perturbation theory (at a perturbed FRW spacetime), but it is not
the case in the non-linear regime. When the axion was initially located
at a plateau region, the instabilities discussed in the previous
sections can lead to a prominent emission of GWs.

\subsection{GW spectrum}
To compute the stochastic background of GWs, we consider the line element with the tensor metric perturbations, $h_{ij}$:
\begin{equation}
ds^2 = -dt^2 + a^2(\delta_{ij} + h_{ij}) dx^i dx^j,
\end{equation}
where we neglect the scalar and vector metric perturbations. Note that $h_{ij}$ satisfies the transverse-traceless condition, $\partial_i h_{ij}=0$ and $h_{ii}=0$. The linearized Einstein equation gives the evolution of the tensor metric perturbation,
\begin{equation} \label{eq:hij}
\ddot{h}_{ij} + 3H\dot{h}_{ij} -\frac{1}{a^2}\nabla^2 h_{ij}= \frac{2}{M_{\rm P}^2} \Pi_{ij}^{\rm TT},
\end{equation}
where $\Pi_{ij}^{\rm TT}$ is the anisotropic stress tensor with the transverse-traceless (TT) projection.
Using the tensor perturbation $h_{ij}$, one can obtain the energy density of the stochastic GW background as follows,
\begin{equation}
\rho_{\rm GW}(t) = \frac{M_{\rm P}^2}{4} \left\langle \dot{h}_{ij} \dot{h}_{ij} \right\rangle.
\end{equation}
Let us redefine the tensor mode as $h_{ij} = \bar{h}_{ij}/a$ and rewrite Eq.\,(\ref{eq:hij}) in terms of the conformal time,
\begin{equation} 
\bar{h}_{ij}''-\partial_{\tilde{\sbm{x}}}^2 \bar{h}_{ij} -\frac{a''}{a}
 \bar{h}_{ij} = 2 a^3 \left(\frac{f}{M_{\rm P}}\right)^2
 \tilde\Pi_{ij}^{\rm TT}, \label{eq:tildehij}
\end{equation}
with $\tilde\Pi_{ij}^{\rm TT} = \Pi_{ij}^{\rm TT}/(mf)^2$.

In our lattice calculation, instead of directly solving Eq.\,(\ref{eq:tildehij}), we solve the following evolution equation of $u_{ij}$,
\begin{equation}
u_{ij}'' -\partial_{\tilde{\sbm{x}}}^2  u_{ij}-\frac{a''}{a}u_{ij} = 2
 a^3 \left(\frac{f}{M_{\rm P}}\right)^2 \tilde\Pi_{ij},  \label{eq:eom_uij}
\end{equation}
where $\tilde\Pi_{ij}$ is the source term before applying the TT projection
which is given by $\tilde\Pi_{ij} = \partial_{\tilde{i}}\tilde\phi \partial_{\tilde{j}} \tilde\phi$. 
One can obtain $\bar{h}_{ij}$ by operating the TT projection on $u_{ij}$ in the Fourier space after solving Eq.\,(\ref{eq:eom_uij}).
In this way, we obtain the same solution as the one obtained by directly solving Eq.\,(\ref{eq:tildehij}) \cite{GarciaBellido:2007af}.

The TT projection can be simply defined in the Fourier space, in which the TT projection operator is given by
\begin{equation}
\Lambda_{ijlm}(\hat{\bf k}) = P_{il}(\hat{\bf k}) P_{jm}(\hat{\bf k})-\frac{1}{2}P_{ij}(\hat{\bf k}) P_{lm}(\hat{\bf k}),
\end{equation}
with $P_{ij} (\hat{\bf k}) = \delta_{ij}-\hat{k}_i \hat{k}_j$ and $\hat{\bf k} = {\bf k}/|{\bf k}|$. Using the projection operator,  we obtain the Fourier mode of the tensor perturbation  as
\begin{equation}
h_{ij}({\bf k}) = \frac{1}{a}\Lambda_{ijlm}(\hat{\bf k}) u_{lm}({\bf k}).
\end{equation}
The energy density of the stochastic GW background can be rewritten in terms of the Fourier transform,
\begin{equation}
\rho_{\rm GW} = \frac{M_{\rm P}^2}{4L^3} \int d^3k \dot{h}_{ij}({\bf k}) \dot{h}^*_{ij}({\bf k}),
\end{equation}
where $L$ is the size of the Universe.
Finally, the spectrum of GW density parameter in terms of the frequency, $\nu$,  can be calculated as
\begin{equation}
\Omega_{\rm GW}(\nu) = \frac{1}{\rho_c} \frac{d\rho_{\rm GW}}{d\ln \nu},
\end{equation}
where $\rho_c$ denotes the total energy density, given by $\rho_c= 3H^2 M_{\rm P}^2$.

\begin{figure}[tp]
\centering
\subfigure[$m\tau=10$--20]{
\includegraphics [width = 7cm, clip]{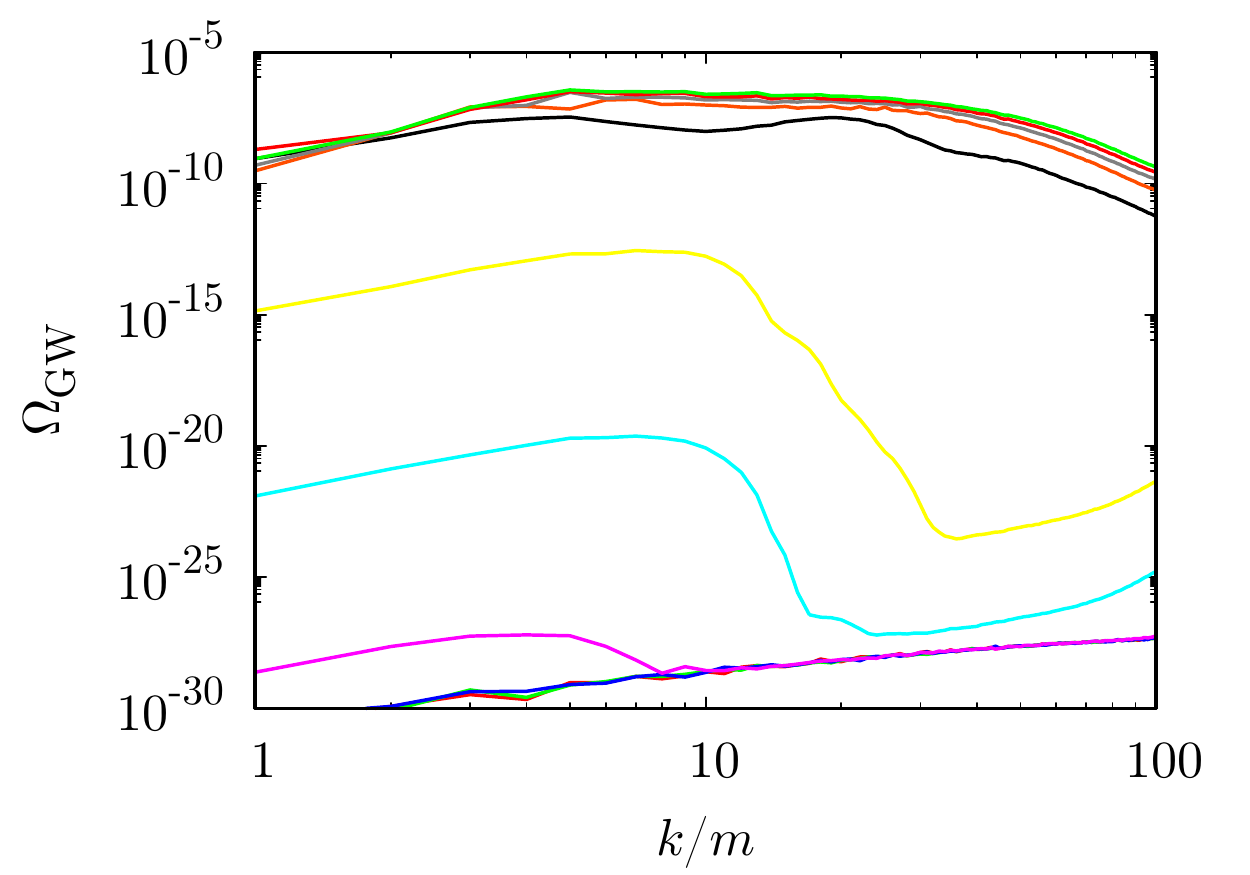}
\label{subfig:OmegaGW1a}
}
\subfigure[$m\tau=20$--75]{
\includegraphics [width = 7cm, clip]{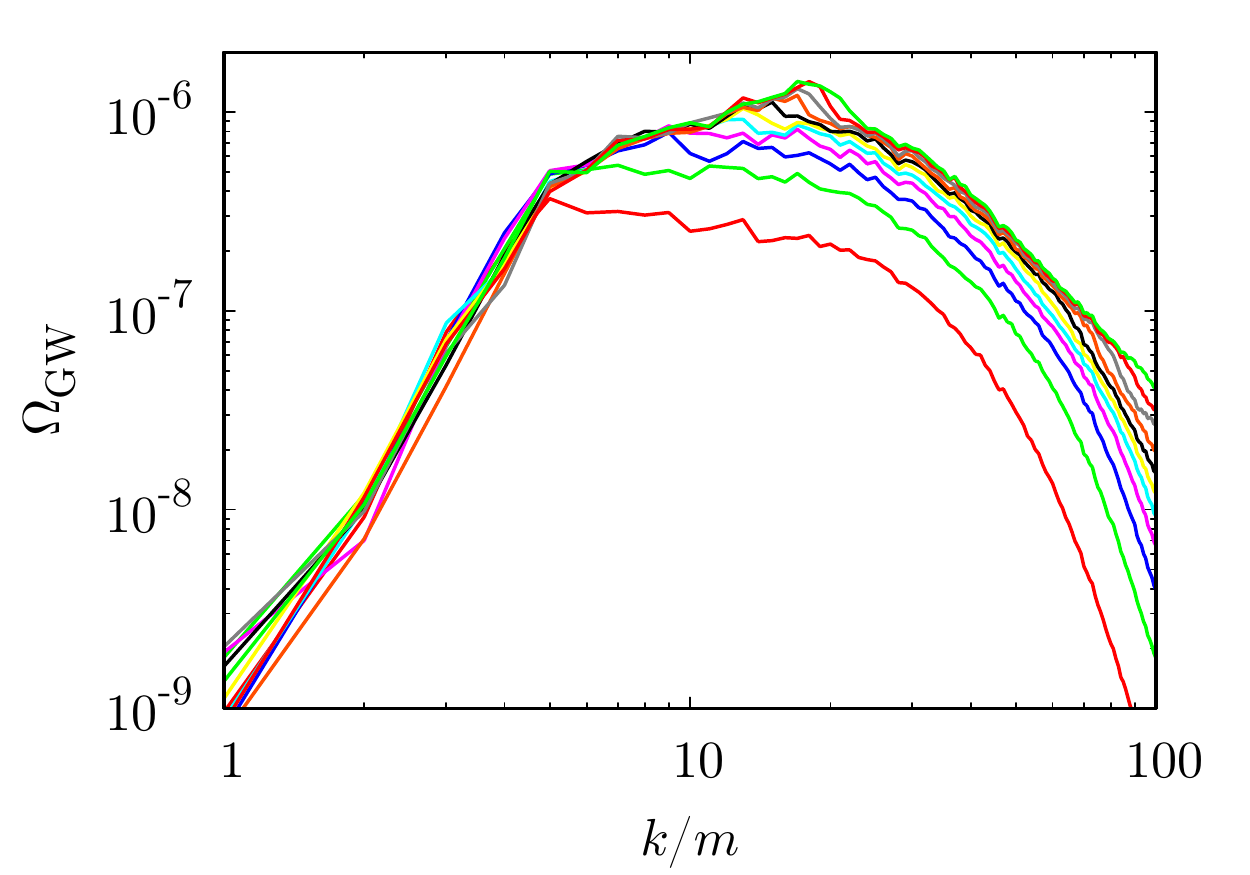}
\label{subfig:OmegaGW1b}
}
\caption{
Evolution of the power spectrum of the density parameter of GW. Time evolves from bottom to top in both panels.
We have taken $f = 10^{16}$ GeV, $c=5$ and $\tilde\phi_i = 3$.
}
\label{fig:OmegaGW1}
\end{figure}

\begin{figure}[tp]
\centering
\subfigure[$m\tau=15$--25]{
\includegraphics [width = 7cm, clip]{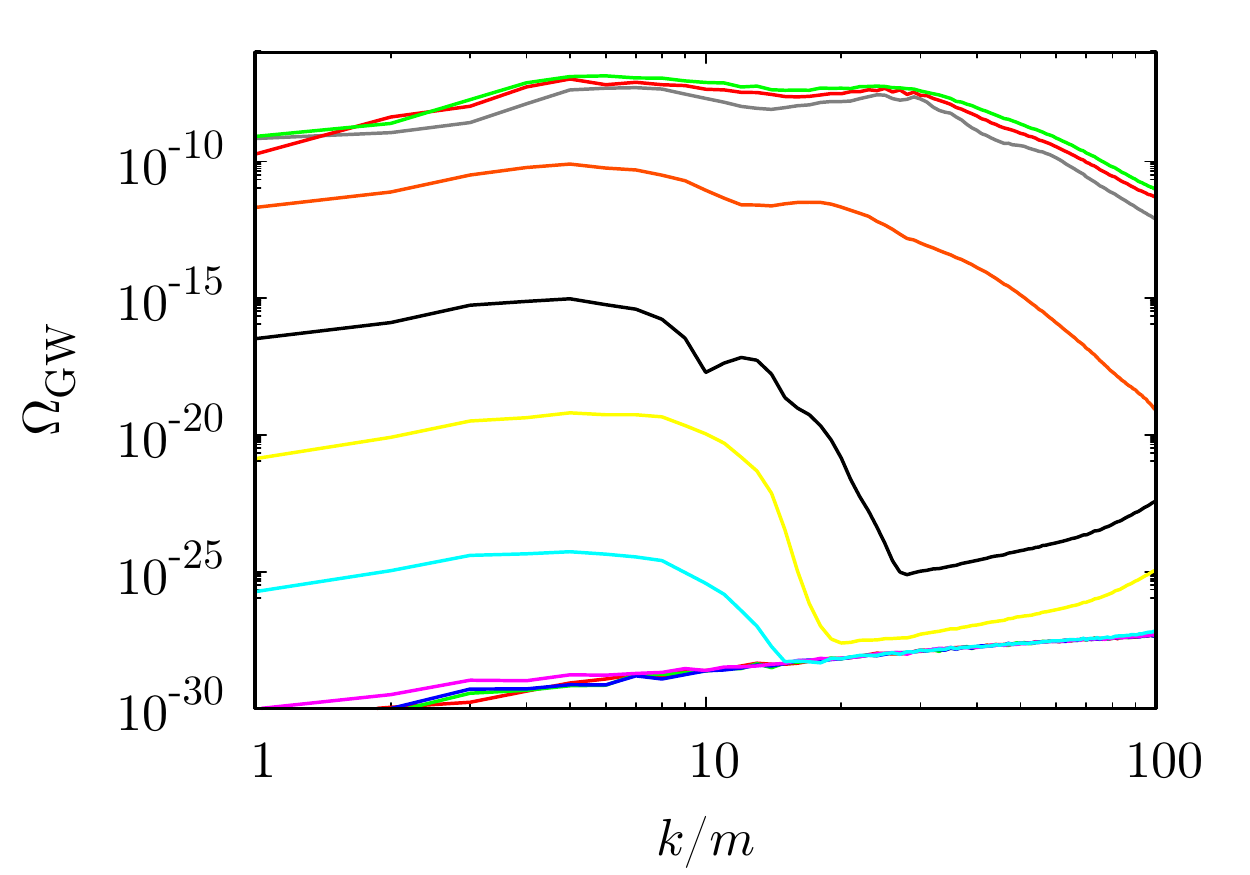}
\label{subfig:OmegaGW2a}
}
\subfigure[$m\tau=25$--75]{
\includegraphics [width = 7cm, clip]{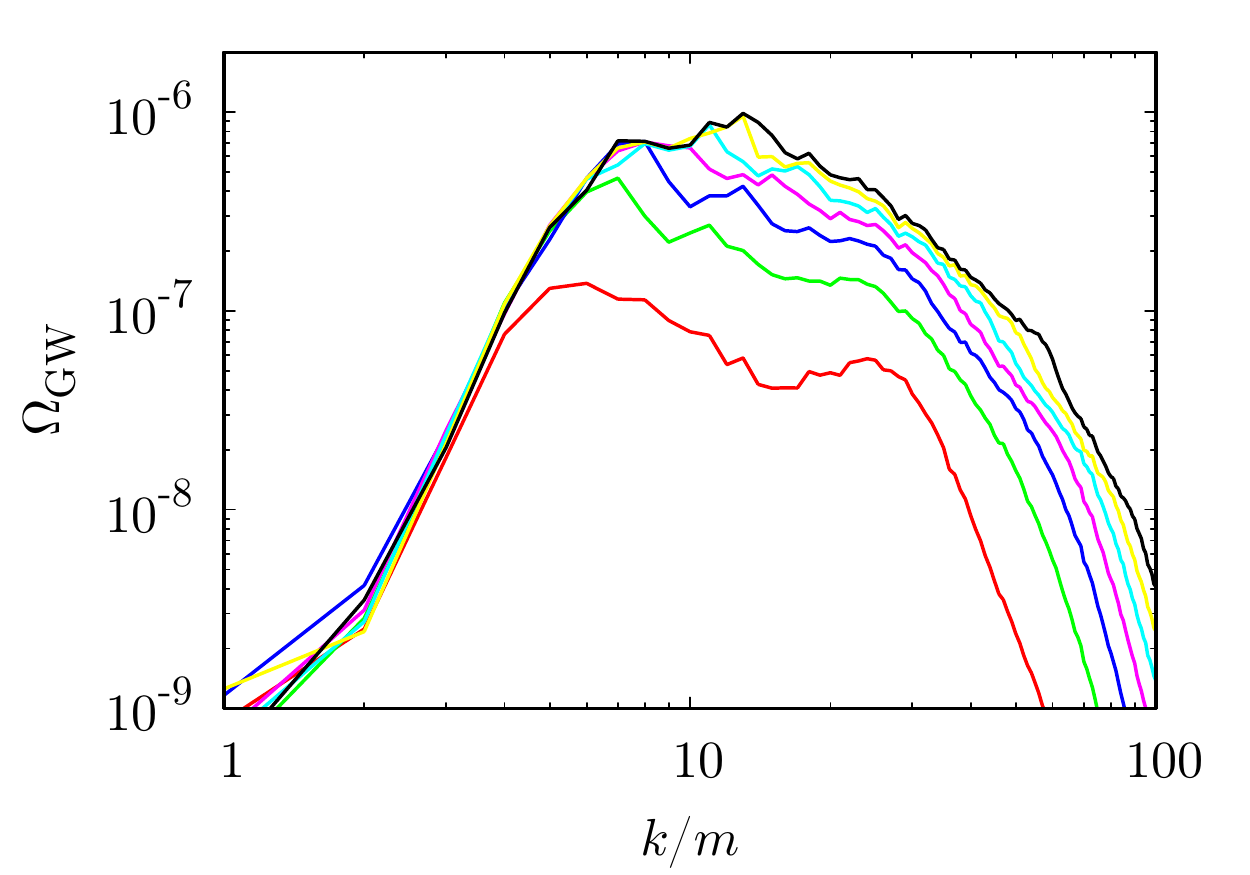}
\label{subfig:OmegaGW2b}
}
\subfigure[$m\tau=10$--20]{
\includegraphics [width = 7cm, clip]{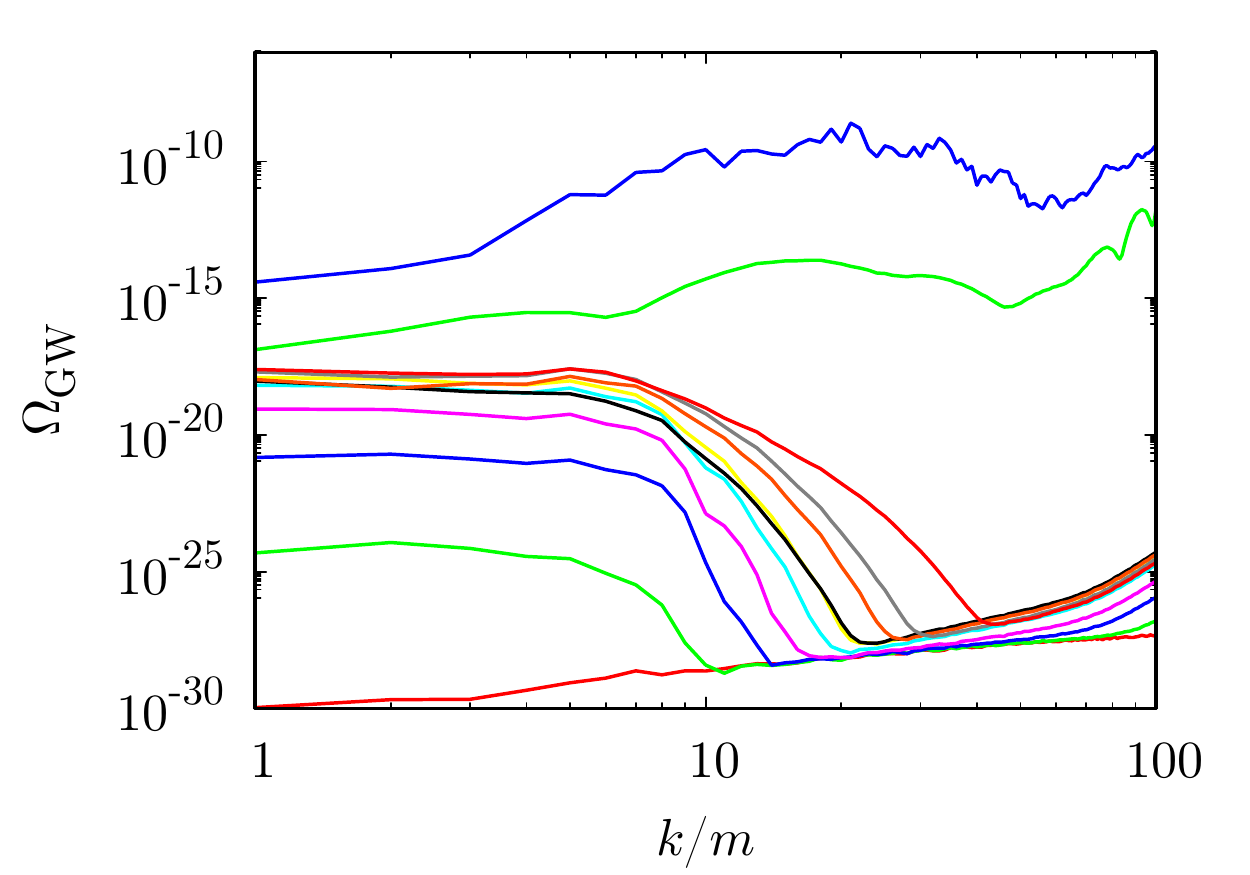}
\label{subfig:OmegaGW3a}
}
\subfigure[$m\tau=20$--75]{
\includegraphics [width = 7cm, clip]{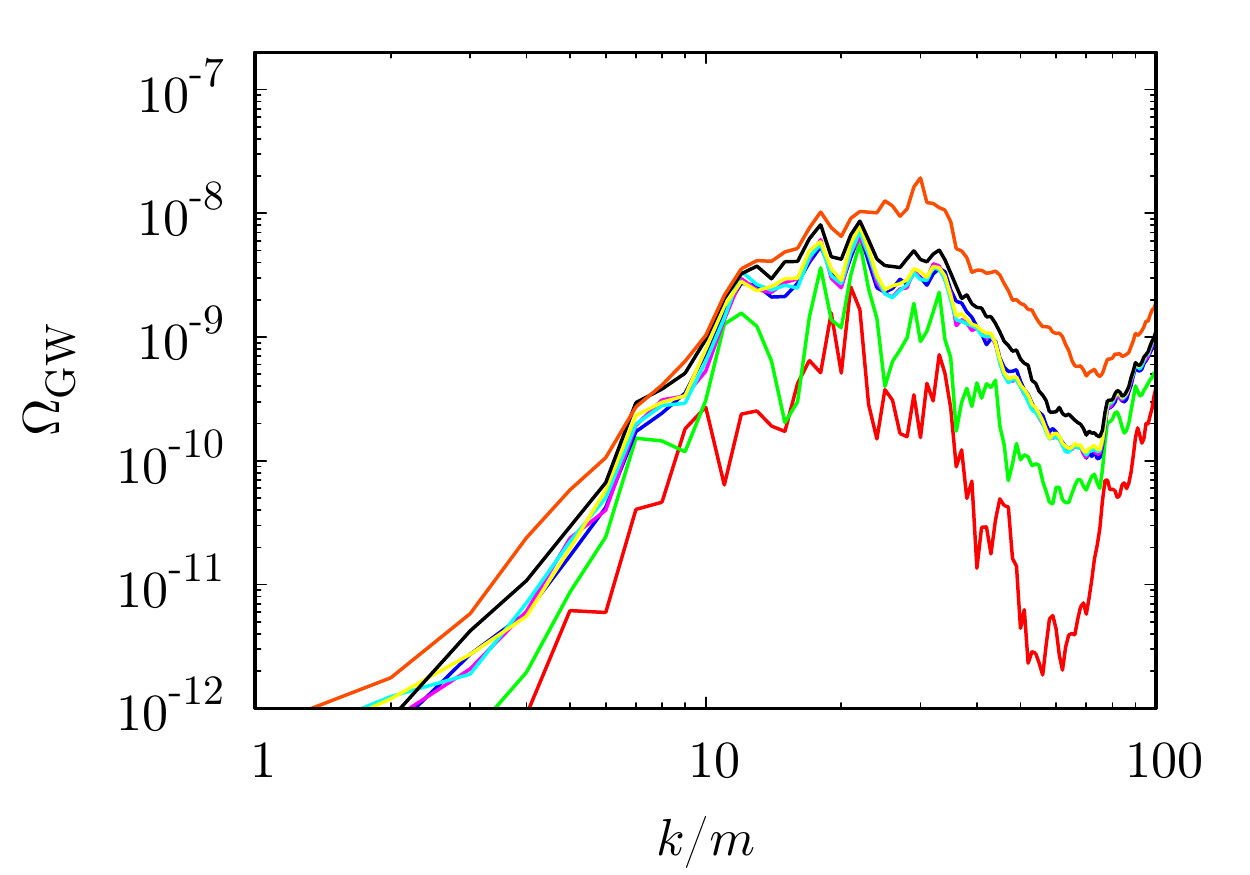}
\label{subfig:OmegaGW3b}
}
\caption{
Same as Fig.~\ref{fig:OmegaGW1} but $c=2$ (top panels) and $c=0$ (bottom panels).
}
\label{fig:OmegaGW23}
\end{figure}

We solved Eq.\,(\ref{eq:eom_uij}) together with Eq.\,(\ref{eq:eom_Phi}) in 3-dimensional lattice space with the same setup in the previous section and calculated the spectrum using the above formula.
Fig.\,\ref{fig:OmegaGW1} and \ref{fig:OmegaGW23} show the evolution of
the spectrum of GW density parameter. They show that for $(c,
\tilde{\phi}_i) = $ (5,\,3) and (2,\,3), the peak wavenumber is around
$k_{\rm peak} \simeq 10m$, which is roughly twice as large as the peak
wave number of the spectrum of $n_k$ before the rescattering.  
We found that the GW emission stops around $m\tau \sim 40$, which
corresponds to the time of oscillon formation as shown in the previous
section. Because the axion field configuration becomes almost
spherically symmetric after the oscillon formation, GWs are no longer
emitted after that \cite{Zhou:2013tsa}.

The GWs emitted from a scalar field which was located at a plateau region has been sometimes described as
``the GWs from oscillons''~\cite{Antusch:2016con,Antusch:2017flz,Antusch:2017vga}. However, this is somehow
misleading, because the prominent GW emission takes place prior to the
formation of the oscillons. In
fact, the efficient GW emission stops after the oscillons formed and the spectrum of the GWs gets
decoupled from the spectrum of the axion number density, where the
momentum transfer still continues. (Related to this, see a recent
publication \cite{Amin:2018xfe}.)

\subsection{GW forest}
So far, we have computed the spectra of the axion and GWs without specifying the
axion mass 
because the evolution equations do not depend explicitly on the axion
mass if we use dimensionless time and spatial coordinates normalized by
the axion mass. However, the onset time of the axion oscillation is determined by the axion mass and thus the current frequency of the
emitted GWs depends on the axion
mass. For our convenience, let us introduce $\kappa$, using the peak
physical frequency $\omega_{\rm phys}$ as 
\begin{align}
  \kappa \equiv \frac{\omega_{\rm phys}}{m} = \frac{k_{\rm peak}^{\rm em}}{m\, a_{\rm em}}\,. 
\end{align}
As discussed in the previous section,
after the rescattering becomes important, the turbulence drives the
momentum flow to UV. Taking into account those, we can express $\kappa$
as
\begin{align}
 & \kappa = \frac{k_{\rm peak}^{\rm res}}{m\, a_{\rm res}}  \times 
\frac{k_{\rm peak}^{\rm em}/a_{\rm em}}{k_{\rm peak}^{\rm res}/a_{\rm res}}\,.
\end{align}
When the dominant instability process is the flapping resonance, 
$k_{\rm peak}^{\rm res}/(m\, a_{\rm res})$ is given by
Eq.~(\ref{Exp:peakFR}) and when it is the narrow resonance, $k_{\rm peak}^{\rm res}/(m\, a_{\rm res})$ 
is given by Eq.~(\ref{Exp:peakNR}). The 
second factor $(k_{\rm peak}^{\rm em}/a_{\rm em})/(k_{\rm peak}^{\rm res}/a_{\rm res})$
describes the momentum flow due to the turbulence.

Using $\kappa$, the redshifted frequency of GWs today is given by
\begin{align}
 & \nu_0 = \frac{\kappa m}{2\pi} \left( \frac{a_{\rm em}}{a_0} \right)\,.
\end{align}
When an axion emitted GWs during radiation domination, we obtain
\begin{equation}
\nu_0 =
 \frac{\kappa m}{2\pi} \times \left( \frac{\rho_{{\rm r}, 0}}{\rho_{\rm r,\,em}}
			      \right)^{1/4}  \simeq 0.78 {\rm nHz}\, \kappa
 \left(\frac{m}{H_{\rm em}} \right)^{1/2}  \left(\frac{m}{10^{-12}{\rm
  eV}}\right)^{1/2} \,,
\end{equation}
where we approximated $\rho_{\rm r,\, em}$ as $\rho_{\rm r,\,em} \simeq \rho_{\rm em}$ and used 
$H_0 = 100 h {\rm km}\, {\rm s}^{-1} {\rm Mpc}^{-1} = 2.13 h \times
10^{-33}$ eV, 
$\Omega_r h^2 \simeq 2.47 \times 10^{-5}$, and $1 {\rm Hz}= 6.58 \times 10^{-16}$eV.
Similarly, when an axion emitted GWs during (late time) matter domination, we obtain
\begin{align}
 & \nu_0 =  \frac{\kappa m}{2\pi} \times \left( \frac{\rho_{{\rm m},
 0}}{\rho_{\rm m,\, em}} \right)^{1/3} \simeq 2.1 \times 10^{-18}
 {\rm Hz}\, \kappa \left( \frac{m}{H_{\rm em}} \right)^{2/3} \left(
 \frac{m}{10^{-30} {\rm eV}} \right)^{\frac{1}{3}} 
\end{align} 
where we approximated $\rho_{\rm m,\, em}$ as 
$\rho_{\rm m,\, em} \simeq \rho_{\rm em}$ and used 
$\Omega_{\rm m} h^2 \simeq 0.14$~\cite{Ade:2015xua}.
%

\begin{figure}[tp]
\centering
\includegraphics [width = 9cm, clip]{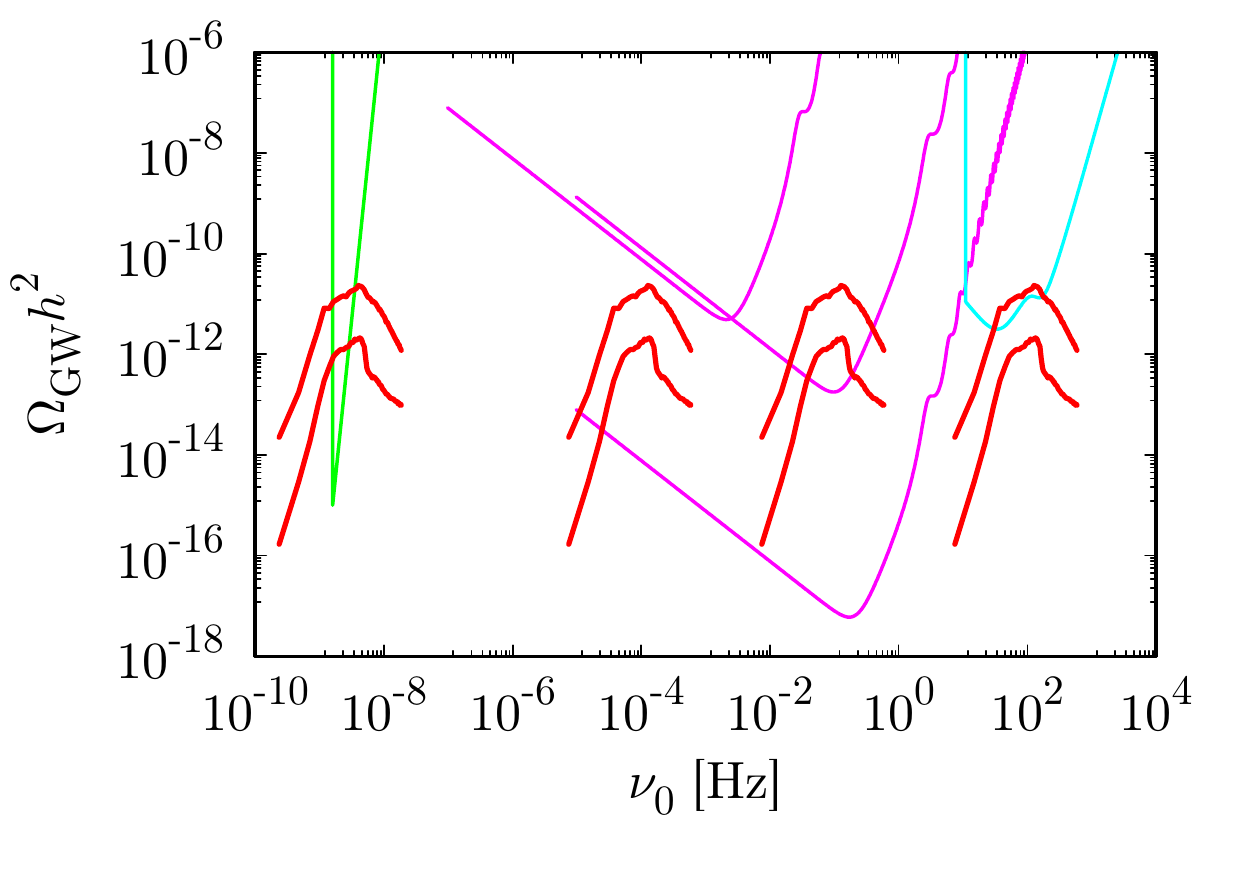}
\caption{
Predicted density spectrum of GWs (thick red lines) and sensitivity
 curves of SKA, LISA, DECIGO (ultimate-DECIGO) and ET from left to
 right. We have taken $f=10^{16}$ GeV, $c=5$ and the axion mass is set
 to be $m=10^{-15}$ eV, $10^{-6}$ eV, 1 eV and $10^6$ eV from left to
 right and $\phi_i=3$ and 2 from top to bottom. 
}
\label{fig:GWforest}
\end{figure}
String theory predicts axions in various mass scales.
When these axions were located at a plateau region before they start to
oscillate, our discussion here predicts GW emissions with
various frequencies, dubbed {\it Gravitational wave forest}.  Figure
\ref{fig:GWforest} shows the density spectrum of GWs from axions
with different mass scales and their detectability by current and future
multi-band GW detectors~\cite{GWground,GWspace}. 
In Fig.~\ref{fig:GWforest}, we chose the decay constant $f$ as $f=10^{16}$
GeV as is typically the case for stringy axions~\cite{Svrcek:2006yi}. When we change $f$,
$\ogw$ scales as $\ogw \propto f^4$.

\subsection{GWs from axion dark matter}
In the previous subsection, we discussed the spectrum of GWs emitted
from axions for a fixed value of $f$. In case axions do not decay into
other species and they are just adiabatically diluted after the emission
of GWs, their abundances should be compatible with the thermal history
of the Universe later on~\cite{review}. 

For our illustrative purpose, here let us consider the case when the
potential of the axion after the onset of the oscillation can be
approximated by a power-law form as $\tilde{V}(\tilde{\phi}) \propto \tilde{\phi}^n$. 
Using this potential, here, we discuss a crude evaluation of $\ogw$
based on an order estimation to understand which quantity
is important for enhancing $\ogw$. A prominent emission of GWs takes
place, roughly when the energy is distributed equally between the
homogeneous mode and the inhomogeneous mode, i.e.,
\begin{align}
 & \left| \frac{\delta \rho_{\rm em}}{\rho_{\rm em} (\langle \phi \rangle)} \right| \simeq {\cal O}(1)\,,
\end{align}
which roughly implies $|\delta \phi| \simeq |\langle \phi \rangle|$. 
Then, using\footnote{For $\tilde{V} \propto \tilde{\phi}^n$, the
equation of state for the homogeneous mode of the axion is
given by $\omega = (n-2)/(n+2)$. Then, the homogeneous modes of
$\tilde{\phi}$ and the energy density $\tilde{\rho}$ scale as $\tilde{\phi} \propto a^{- 6/(n+2)}$ and $\tilde\rho \propto a^{- 6n/(n+2)}$. } 
$|\langle \phi_{\rm em} \rangle| \simeq  f (a_{\rm osc}/a_{\rm em})^{6n/(n+2)}$, 
we obtain the amplitude of the GWs emitted at $a \simeq a_{\rm em}$ as (we drop the tensor indices),
\begin{align}
 & |h_{\rm em}| \simeq 2   \Delta  \left( \frac{f}{M_P} \right)^{\!2}
 \left( a_{\rm osc} \over a_{\rm em} \right)^{12/(n+2)} \,, \label{Exp:orderh}
\end{align}
where $\Delta$ quantifies the efficiency of the GW emission from the
inhomogeneous mode of the axion and takes a value in the range $0 \leq \Delta < 1$. 
In particular, for a GW emission after formations of oscillons, which
are almost spherically symmetric, $|h_{\rm em}|$ is highly suppressed by
$\Delta \ll 1$.

Using Eq.~(\ref{Exp:orderh}), the energy density of GWs at the peak wavenumber is given by
\begin{align}
 & \rho_{\rm GW, em} \simeq 
 \frac{M_P^2\, \omega_{\rm phys}^2}{4}  (h_{\rm em})^2  \simeq \Delta^2 (\kappa m)^2
 \frac{f^4}{M_{pl}^2}  \left( a_{\rm osc} \over a_{\rm em} \right)^{24/(n+2)}  \,.  \label{Exp:rhoGWrs}
\end{align}
Dividing this expression by the energy density of radiation at 
$a=a_{\rm em}$, we obtain
\begin{align}
 &  \frac{\rho_{{\rm GW},0}}{\rho_{r, 0}}\simeq \frac{\rho_{\rm
 GW,\, em}}{\rho_{r, {\rm em}}} \simeq \frac{\kappa^4
 \Delta^2}{M_P^2} \frac{1}{(2\pi \nu_0)^2} \frac{(\rho_{\rm
 em})^2}{\rho_{r,{\rm em}}} \left( \frac{a_{\rm em}}{a_0} \right)^2
 \left( \frac{a_{\rm osc}}{a_{\rm em}} \right)^{\frac{12(2-n)}{2 +
 n}}\,, \label{Exp:oest}
\end{align}
where we used $\rho_{\rm em} \simeq  \rho_{\rm osc} (a_{\rm osc} / a_{\rm em})^{6n/(n+2)}$.
In particular when the phase 2 continues long, we cannot express the
potential in terms of a single power low term $\tilde{\phi}^n$. However,
Eq.~(\ref{Exp:oest}) is somewhat instructive. For $|\tilde{\phi}| \agt 1$, the potential $\tilde{V}$ is shallower than $\tilde{\phi}^2$.
Equation (\ref{Exp:oest}) tells us that the emitted GWs are more
suppressed for $n<2$, when it takes longer until the emission of the GWs
after the onset of the oscillation. This suppression can be evaded,
either in case $a_{\rm osc} \simeq a_{\rm em}$ or 
(even if $a_{\rm em} \gg a_{\rm osc}$) in case the phase 2 finishes
soon and the dominant instability process takes place during the phase
3.

For a simplistic estimation, here let us consider the case when we can
approximate the potential as $\tilde{V} \simeq \tilde{\phi}^2/2$ soon
after $\tilde{\phi}$ starts to oscillate. (We end up with the same estimation, when
$a_{\rm osc} \simeq a_{\rm em}$ and $\tilde{V}$ can be approximated as
the quadratic one after the GW emission.) Then, multiplying $\Omega_r h^2$ 
on Eq.~(\ref{Exp:oest}) with $n=2$, we obtain
\begin{align}
  \ogw h^2 \simeq
 \frac{3 \kappa^4 \Delta^2}{(2 \pi \nu_0)^2} \left( \frac{H_0}{h}
 \right)^2 (\Omega_\phi h^2)^2 \simeq 0.8 \times 10^{-18} \kappa^4\Delta^2 \left(
 \frac{{\rm nHz}}{\nu_0} \right)^2  (\Omega_\phi h^2)^2\,. 
\end{align}
Since $\Delta < 1$ and $\Omega_\phi h^2 \leq \Omega_{\rm CDM}h^2 \simeq 0.12$, this estimation reads
\begin{align}
 & \ogw h^2 < 1.6 \times 10^{-16} \left( \frac{\kappa}{10} \right)^4 \left(
 \frac{{\rm nHz}}{\nu_0} \right)^2\,,
\end{align}
indicating that to reach $\ogw h^2 \sim 10^{-16}$ at
$\nu_0=$nHz, $\kappa$ should be larger than 10. At lower frequencies
than nHz, $\ogw$ is enhanced as $ \propto 1/\nu^2_0$. To detect GWs in this
frequency range, we will need a new window of GW detections which
fills the gap between CMB and PTA observations~\cite{PTA}. 
\begin{figure}[tp]
\centering
\includegraphics [width = 9cm, clip]{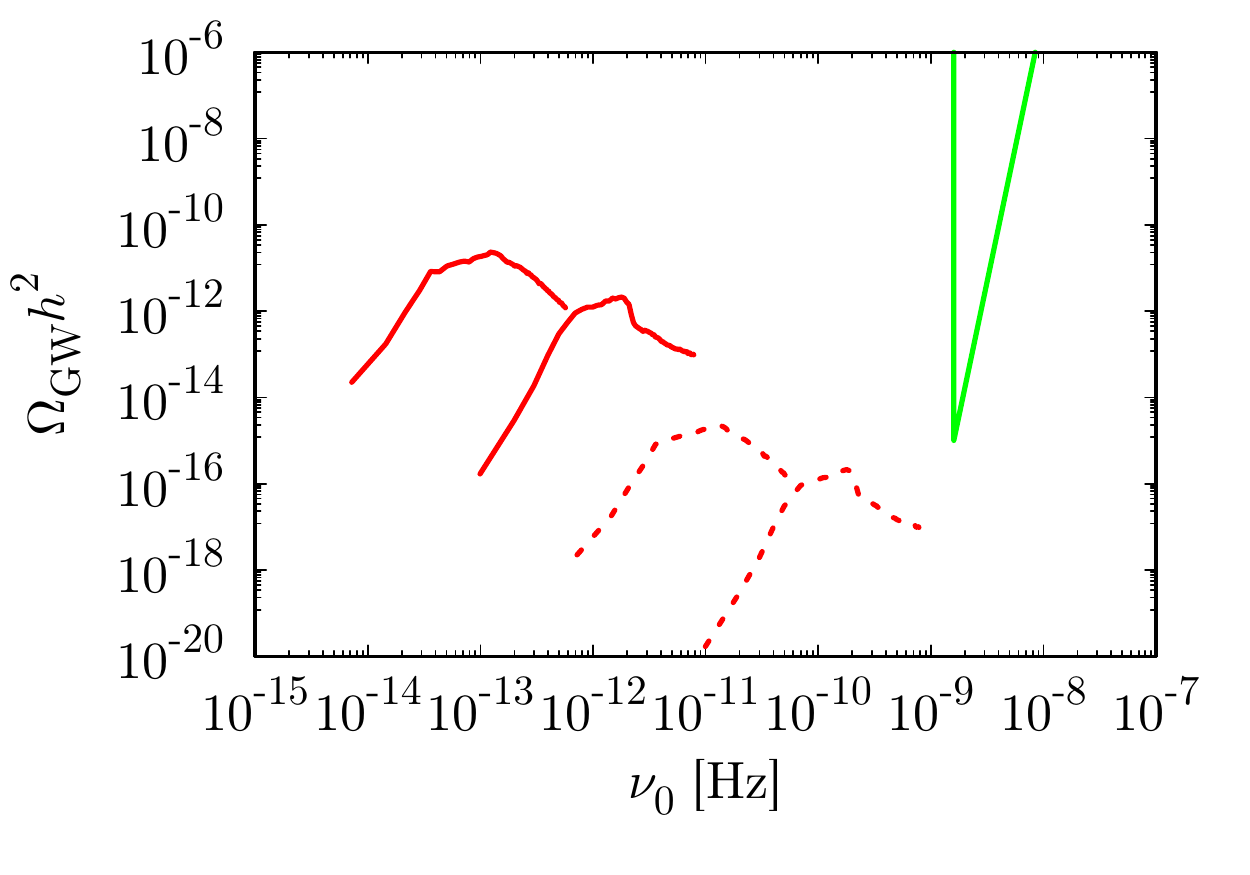}
\caption{Same as Fig.~\ref{fig:GWforest} but the present abundance of
 the axion is fixed to be the dark matter abundance. The dashed lines
 correspond to $f=10^{15}$ GeV.  
}
\label{fig:GWforest2}
\end{figure}
Figure \ref{fig:GWforest2} shows $\ogw h^2$ computed from the lattice simulation for the $\alpha$ attractor
potential with $c=5$ and $\tilde{\phi}_i= 2,\, 3$, when $\Omega_\phi = \Omega_{\rm CDM}$ and $n=2$.

\section{Conclusion}
In this paper, we analyzed the dynamics of string axions which were
initially located at a plateau region in their scalar potentials. 
After the delayed onset of the axion oscillation, different instabilities take
place, depending on different phases which are characterized by the signature of
$\tilde{V}_{\tilde{\phi} \tilde{\phi}}$. In particular, when the
tahyonic instability repeatedly turns on and off, the field
fluctuation is resonantly enhanced (flapping
resonance). The flapping resonance can more efficiently amplify the
field fluctuation than the narrow resonance. In addition,
when the plateau condition, the condition ii), is fulfilled, oscillons
are rather generically formed as a consequence of resonant
instabilities. This was also verified in our lattice simulation. 
Such a resonant amplification of the axion fluctuation can lead to a
copious emission of GWs, providing a new source of gravitational wave
background. The GW emission lasts also after the saturation of the
exponential growth but terminates after the oscillon formation.

In the context of string axiverse, there are plenty of axions whose mass
spectrum spreads over many orders of magnitude. When some of the axions
were located at plateau regions before they start to oscillate, significant
amounts of GWs can be emitted in various frequency bands,
corresponding to their mass scales  ({\it GW forest}). Axions with $f \sim 10^{16}$GeV, which are typically predicted in
string theory, lead to detectable GWs. This can open up a new window to probe string axiverse by
means of future multi-band GW observations.

One caveat is that string axions with $f \sim 10^{16}$GeV overclose the Universe, when they
are long-lived. This problem can be circumvented, if these axions decay,
e.g., due to an enhanced coupling with gauge fields.
It occurs e.g. in clockwork/aligned axion model \cite{Agrawal:2017cmd}
and the model with gauge kinetic mixing \cite{Daido:2018dmu}. In
addition, a moderately large coupling to (hidden) gauge fields can
suppress the final axion abundance through efficient dissipation of the
axion energy into gauge field
\cite{Agrawal:2017eqm,Kitajima:2017peg}. It potentially opens the
possibility to generate GWs in the reach of pulsar timing observation.

A sizable coupling between axions and gauge fields can alter the GW
emission process. In this case, the axion
can produce gauge fields explosively and the backreaction from the
produced gauge fields yields efficient production of the nonzero mode
axions \cite{Kitajima:2017peg}. In that case, both axion and gauge field
can source GWs \cite{Adshead:2018doq}. In particular, GW emission
continues even after the oscillon formation because of the efficient
gauge field production inside the oscillon and it predicts GWs with
higher frequency range. We will report this result in our
future study~\cite{inprep}.

\acknowledgments
We would like to thank K.~Kamada and M.~Yamazaki for helpful
discussions. N.~K. acknowledges the support by Grant-in-Aid for JSPS
Fellows. J.~S. was in part supported by JSPS KAKENHI
Grant Numbers JP17H02894, JP17K18778, JP15H05895, JP17H06359, JP18H04589.
J.~S and Y.~U. are also supported by JSPS Bilateral Joint Research
Projects (JSPS-NRF collaboration) “String Axion Cosmology.”
Y.~U. is supported by JSPS Grant-in-Aid for Research Activity Start-up
under Contract No.~26887018, Grant-in-Aid for Scientific Research on Innovative Areas under
Contract Nos.~16H01095 and 18H04349, and Grant-in-Aid for Young Scientists (B) under
Contract No.~16K17689. Y.~U. is also supported in part by Building of
Consortia for the Development of Human Resources in 
Science and Technology and Daiko Foundation.



\begin{thebibliography}{99}

\bibitem{Abbott:2016blz} 
  B.~P.~Abbott {\it et al.} [LIGO Scientific and Virgo Collaborations],
  Phys.\ Rev.\ Lett.\  {\bf 116}, no. 6, 061102 (2016)
  doi:10.1103/PhysRevLett.116.061102
  [arXiv:1602.03837 [gr-qc]].


\bibitem{Ade:2017uvt} 
  P.~A.~R.~Ade {\it et al.} [POLARBEAR Collaboration],
  Astrophys.\ J.\  {\bf 848}, no. 2, 121 (2017)
  doi:10.3847/1538-4357/aa8e9f
  [arXiv:1705.02907 [astro-ph.CO]].


\bibitem{Matsumura:2013aja} 
  T.~Matsumura {\it et al.},
  J.\ Low.\ Temp.\ Phys.\  {\bf 176}, 733 (2014)
  doi:10.1007/s10909-013-0996-1
  [arXiv:1311.2847 [astro-ph.IM]].


\bibitem{GWspace} 
  P.~Amaro-Seoane {\it et al.},
  GW Notes {\bf 6}, 4 (2013).
  N.~Seto, S.~Kawamura and T.~Nakamura,
  Phys.\ Rev.\ Lett.\  {\bf 87}, 221103 (2001)


\bibitem{GWground} 
  J.~Aasi {\it et al.} [LIGO Scientific Collaboration],
  Class.\ Quant.\ Grav.\  {\bf 32}, 074001 (2015). 
  F.~Acernese {\it et al.} [VIRGO Collaboration],
  Class.\ Quant.\ Grav.\  {\bf 32}, no. 2, 024001 (2015).
   K.~Somiya [KAGRA Collaboration],
  Class.\ Quant.\ Grav.\  {\bf 29}, 124007 (2012)
  

  
\bibitem{PTA} 
  S.~L.~Detweiler,
  ``Pulsar timing measurements and the search for gravitational waves,''
  Astrophys.\ J.\  {\bf 234}, 1100 (1979).
  L.~Lentati {\it et al.},
  ``European Pulsar Timing Array Limits On An Isotropic Stochastic Gravitational-Wave Background,''
  Mon.\ Not.\ Roy.\ Astron.\ Soc.\  {\bf 453}, no. 3, 2576 (2015).
  [arXiv:1504.03692 [astro-ph.CO]].
  Z.~Arzoumanian {\it et al.} [NANOGrav Collaboration],
  ``The NANOGrav Nine-year Data Set: Limits on the Isotropic Stochastic Gravitational Wave Background,''
  Astrophys.\ J.\  {\bf 821}, no. 1, 13 (2016).
  [arXiv:1508.03024 [astro-ph.GA]].


  
  
\bibitem{Svrcek:2006yi} 
  P.~Svrcek and E.~Witten,
  JHEP {\bf 0606}, 051 (2006)
  doi:10.1088/1126-6708/2006/06/051
  [hep-th/0605206].

\bibitem{LVS}
  J.~P.~Conlon, F.~Quevedo and K.~Suruliz,
  JHEP {\bf 0508}, 007 (2005). J.~Halverson, C.~Long and P.~Nath,
  arXiv:1703.07779 [hep-ph].




\bibitem{StAx}
  A.~Arvanitaki, S.~Dimopoulos, S.~Dubovsky, N.~Kaloper and J.~March-Russell,
  Phys.\ Rev.\ D {\bf 81}, 123530 (2010).



\bibitem{JY17} 
  J.~Soda and Y.~Urakawa,
  arXiv:1710.00305 [astro-ph.CO].

\bibitem{Dubovsky:2011tu} 
  S.~Dubovsky, A.~Lawrence and M.~M.~Roberts,
  JHEP {\bf 1202}, 053 (2012)


\bibitem{NWY} 
  Y.~Nomura, T.~Watari and M.~Yamazaki,
  arXiv:1706.08522 [hep-ph].




\bibitem{KLS_L} 
  L.~Kofman, A.~D.~Linde and A.~A.~Starobinsky,
  Phys.\ Rev.\ Lett.\  {\bf 73}, 3195 (1994)
  doi:10.1103/PhysRevLett.73.3195
  [hep-th/9405187].

\bibitem{KLS} 
  L.~Kofman, A.~D.~Linde and A.~A.~Starobinsky,
  Phys.\ Rev.\ D {\bf 56}, 3258 (1997)
  doi:10.1103/PhysRevD.56.3258
  [hep-ph/9704452].



\bibitem{KT97} 
  S.~Y.~Khlebnikov and I.~I.~Tkachev,
  Phys.\ Rev.\ D {\bf 56}, 653 (1997)
  doi:10.1103/PhysRevD.56.653
  [hep-ph/9701423].

\bibitem{Dani17} 
  D.~G.~Figueroa and F.~Torrenti,
  JCAP {\bf 1710}, no. 10, 057 (2017)
  doi:10.1088/1475-7516/2017/10/057
  [arXiv:1707.04533 [astro-ph.CO]].



\bibitem{GarciaBellido:2007af} 
  J.~Garcia-Bellido, D.~G.~Figueroa and A.~Sastre,
  Phys.\ Rev.\ D {\bf 77}, 043517 (2008)
  doi:10.1103/PhysRevD.77.043517
  [arXiv:0707.0839 [hep-ph]].


\bibitem{Witten:1979vv} 
  E.~Witten,
  Nucl.\ Phys.\ B {\bf 156}, 269 (1979).

\bibitem{Witten:1980sp} 
  E.~Witten,
  Annals Phys.\  {\bf 128}, 363 (1980).

 

\bibitem{Yonekura:2014oja} 
  K.~Yonekura,
  JCAP {\bf 1410}, no. 10, 054 (2014)
  doi:10.1088/1475-7516/2014/10/054
  [arXiv:1405.0734 [hep-th]].




\bibitem{NY17} 
  Y.~Nomura and M.~Yamazaki,
  Phys.\ Lett.\ B {\bf 780}, 106 (2018)
  doi:10.1016/j.physletb.2018.02.071
  [arXiv:1711.10490 [hep-ph]].


\bibitem{Planck15} 
  P.~A.~R.~Ade {\it et al.} [Planck Collaboration],
  Astron.\ Astrophys.\  {\bf 594}, A20 (2016)
  doi:10.1051/0004-6361/201525898
  [arXiv:1502.02114 [astro-ph.CO]].

 

\bibitem{Shiu:2018wzf} 
  G.~Shiu and W.~Staessens,
  arXiv:1807.00620 [hep-th].

\bibitem{Shiu:2018unx} 
  G.~Shiu and W.~Staessens,
  arXiv:1807.00888 [hep-th].

\bibitem{Czerny:2014wza} 
  M.~Czerny and F.~Takahashi,
  Phys.\ Lett.\ B {\bf 733}, 241 (2014)
  doi:10.1016/j.physletb.2014.04.039
  [arXiv:1401.5212 [hep-ph]].

\bibitem{Czerny:2014xja} 
  M.~Czerny, T.~Higaki and F.~Takahashi,
  JHEP {\bf 1405}, 144 (2014)
  doi:10.1007/JHEP05(2014)144
  [arXiv:1403.0410 [hep-ph]].

\bibitem{Kallosh:2013hoa} 
  R.~Kallosh and A.~Linde,
  JCAP {\bf 1307}, 002 (2013)
  doi:10.1088/1475-7516/2013/07/002
  [arXiv:1306.5220 [hep-th]].

\bibitem{Kallosh:2013yoa} 
  R.~Kallosh, A.~Linde and D.~Roest,
  JHEP {\bf 1311}, 198 (2013)
  doi:10.1007/JHEP11(2013)198
  [arXiv:1311.0472 [hep-th]]. 

\bibitem{Kallosh:2013tua} 
  R.~Kallosh, A.~Linde and D.~Roest,
  Phys.\ Rev.\ Lett.\  {\bf 112}, no. 1, 011303 (2014)
  doi:10.1103/PhysRevLett.112.011303
  [arXiv:1310.3950 [hep-th]].

\bibitem{Ferrara:2013rsa} 
  S.~Ferrara, R.~Kallosh, A.~Linde and M.~Porrati,
  Phys.\ Rev.\ D {\bf 88}, no. 8, 085038 (2013)
  doi:10.1103/PhysRevD.88.085038
  [arXiv:1307.7696 [hep-th]].



\bibitem{Amin17} 
  K.~D.~Lozanov and M.~A.~Amin,
  arXiv:1710.06851 [astro-ph.CO].



\bibitem{Zhang:2017flu} 
  U.~H.~Zhang and T.~Chiueh,
  Phys.\ Rev.\ D {\bf 96}, no. 2, 023507 (2017)
  doi:10.1103/PhysRevD.96.023507
  [arXiv:1702.07065 [astro-ph.CO]].


\bibitem{Zhang:2017dpp} 
  U.~H.~Zhang and T.~Chiueh,
  Phys.\ Rev.\ D {\bf 96}, no. 6, 063522 (2017)
  doi:10.1103/PhysRevD.96.063522
  [arXiv:1705.01439 [astro-ph.CO]].






\bibitem{Felder:2000hj} 
  G.~N.~Felder, J.~Garcia-Bellido, P.~B.~Greene, L.~Kofman, A.~D.~Linde and I.~Tkachev,
  Phys.\ Rev.\ Lett.\  {\bf 87}, 011601 (2001)
  doi:10.1103/PhysRevLett.87.011601
  [hep-ph/0012142].


\bibitem{Antusch:2015nla} 
  S.~Antusch, D.~Nolde and S.~Orani,
  JCAP {\bf 1506}, no. 06, 009 (2015)
  doi:10.1088/1475-7516/2015/06/009
  [arXiv:1503.06075 [hep-ph]].


\bibitem{Brax:2010ai} 
  P.~Brax, J.~F.~Dufaux and S.~Mariadassou,
  Phys.\ Rev.\ D {\bf 83}, 103510 (2011)
  doi:10.1103/PhysRevD.83.103510
  [arXiv:1012.4656 [hep-th]].






\bibitem{MT02} 
  R.~Micha and I.~I.~Tkachev,
  Phys.\ Rev.\ Lett.\  {\bf 90}, 121301 (2003)
  doi:10.1103/PhysRevLett.90.121301
  [hep-ph/0210202].


\bibitem{MT04} 
  R.~Micha and I.~I.~Tkachev,
  Phys.\ Rev.\ D {\bf 70}, 043538 (2004)
  doi:10.1103/PhysRevD.70.043538
  [hep-ph/0403101].





\bibitem{oscillon}
  S.~Kasuya, M.~Kawasaki and F.~Takahashi,
  Phys.\ Lett.\ B {\bf 559}, 99 (2003),
  M.~A.~Amin and D.~Shirokoff,
  Phys.\ Rev.\ D {\bf 81}, 085045 (2010).
  M.~A.~Amin, R.~Easther, H.~Finkel, R.~Flauger and M.~P.~Hertzberg,
  Phys.\ Rev.\ Lett.\  {\bf 108}, 241302 (2012). M.~A.~Amin, M.~P.~Hertzberg, D.~I.~Kaiser and J.~Karouby,
  Int.\ J.\ Mod.\ Phys.\ D {\bf 24}, 1530003 (2014).



\bibitem{Zhou:2013tsa} 
  S.~Y.~Zhou, E.~J.~Copeland, R.~Easther, H.~Finkel, Z.~G.~Mou and P.~M.~Saffin,
  JHEP {\bf 1310}, 026 (2013)



\bibitem{Antusch:2016con} 
  S.~Antusch, F.~Cefala and S.~Orani,
  Phys.\ Rev.\ Lett.\  {\bf 118}, no. 1, 011303 (2017)
  
\bibitem{Antusch:2017flz} 
  S.~Antusch, F.~Cefala, S.~Krippendorf, F.~Muia, S.~Orani and F.~Quevedo,
  arXiv:1708.08922 [hep-th].
  
 \bibitem{Antusch:2017vga} 
  S.~Antusch, F.~Cefala and S.~Orani,
  arXiv:1712.03231 [astro-ph.CO].

 
\bibitem{Amin:2018xfe} 
  M.~A.~Amin, J.~Braden, E.~J.~Copeland, J.~T.~Giblin, C.~Solorio, Z.~J.~Weiner and S.~Y.~Zhou,
  arXiv:1803.08047 [astro-ph.CO].



\bibitem{Ade:2015xua} 
  P.~A.~R.~Ade {\it et al.} [Planck Collaboration],
  Astron.\ Astrophys.\  {\bf 594}, A13 (2016)
  doi:10.1051/0004-6361/201525830
  [arXiv:1502.01589 [astro-ph.CO]].





\bibitem{review} 
  D.~J.~E.~Marsh,
  Phys.\ Rept.\  {\bf 643}, 1 (2016). L.~Hui, J.~P.~Ostriker, S.~Tremaine and E.~Witten,
  Phys.\ Rev.\ D {\bf 95}, no. 4, 043541 (2017).
  






  

  

\bibitem{Agrawal:2017cmd} 
  P.~Agrawal, J.~Fan, M.~Reece and L.~T.~Wang,
  JHEP {\bf 1802}, 006 (2018)
  doi:10.1007/JHEP02(2018)006
  [arXiv:1709.06085 [hep-ph]].



\bibitem{Daido:2018dmu} 
  R.~Daido, F.~Takahashi and N.~Yokozaki,
  Phys.\ Lett.\ B {\bf 780}, 538 (2018)
  doi:10.1016/j.physletb.2018.03.039
  [arXiv:1801.10344 [hep-ph]].
  
\bibitem{Agrawal:2017eqm} 
  P.~Agrawal, G.~Marques-Tavares and W.~Xue,
  JHEP {\bf 1803}, 049 (2018)
  doi:10.1007/JHEP03(2018)049
  [arXiv:1708.05008 [hep-ph]].
    
\bibitem{Kitajima:2017peg} 
  N.~Kitajima, T.~Sekiguchi and F.~Takahashi,
  Phys.\ Lett.\ B {\bf 781}, 684 (2018)
  doi:10.1016/j.physletb.2018.04.024
  [arXiv:1711.06590 [hep-ph]].


\bibitem{Adshead:2018doq} 
  P.~Adshead, J.~T.~Giblin and Z.~J.~Weiner,
  arXiv:1805.04550 [astro-ph.CO].

\bibitem{inprep}
N.~Kitajima, J.~Soda and Y.~Urakawa in preparation.



\end{thebibliography}
\end{document}